\documentclass[12pt]{iopart}
\usepackage{iopams}
\usepackage{bm}
\usepackage{dsfont}
\usepackage{graphicx}
\usepackage{setstack}
\usepackage{mathrsfs} 
\usepackage{xcolor}
\usepackage{braket}
\usepackage[colorlinks=true,breaklinks=false,allcolors=blue]{hyperref}

\usepackage{etoolbox}

\usepackage{lipsum}

\newcommand\dd{{\mathrm{d}}}
\newcommand\ee{{\mathrm{e}}}
\newcommand\ii{{\mathrm{i}}}
\newcommand\dist{{\mathit{d}}}

\begin{document}

\title[Soundcone and Supersonic Propagation in Models with Power Law Interactions]{Interplay of Soundcone and Supersonic Propagation in Lattice Models with Power Law Interactions}  
\author{David-Maximilian Storch$^1$, Mauritz van den Worm$^{2,3}$ and Michael Kastner$^{2,3}$}
\address{$^{1}$ Department of Physics, Georg-August-Universit\"at G\"ottingen, Friedrich-Hund-Platz~1, 37077 G\"ottingen, Germany}
\address{$^{2}$ National Institute for Theoretical Physics (NITheP), Stellenbosch 7600, South~Africa}
\address{$^{3}$ Institute of Theoretical Physics,  University of Stellenbosch, Stellenbosch 7600, South~Africa}
\ead{kastner@sun.ac.za}

\begin{abstract}
We study the spreading of correlations and other physical quantities in quantum lattice models with interactions or hopping decaying like $r^{-\alpha}$ with the distance $r$. Our focus is on exponents $\alpha$ between 0 and 6, where the interplay of long- and short-range features gives rise to a complex phenomenology and interesting physical effects, and which is also the relevant range for experimental realizations with cold atoms, ions, or molecules. We present analytical and numerical results, providing a comprehensive picture of spatio-temporal propagation. Lieb-Robinson-type bounds are extended to strongly long-range interactions where $\alpha$ is smaller than the lattice dimension, and we report particularly sharp bounds that are capable of reproducing regimes with soundcone as well as supersonic dynamics. Complementary lower bounds prove that faster-than-soundcone propagation occurs for $\alpha<2$ in any spatial dimension, although cone-like features are shown to also occur in that regime. Our results provide guidance for optimizing experimental efforts to harness long-range interactions in a variety of quantum information and signaling tasks.
\end{abstract}




\section{Introduction}

Traditionally, the study of lattice models has focused on Hamiltonians where interactions and/or hopping is restricted to a few neighboring sites. Only recently there has been a surge of interest in long-range interacting systems where interaction strengths or hopping amplitudes decay like a power law $r^{-\alpha}$ at large distances $r$. This interest was triggered on the experimental side by progress in the control of ultra-cold atoms, molecules, and ions, which led to the realization of a variety of long-range systems. Examples include magnetic atoms \cite{dePaz_etal13}, polar molecules \cite{Yan_etal13}, trapped ions \cite{Britton_etal12,Islam_etal13,Jurcevic_etal14,Richerme_etal14}, Rydberg atoms \cite{Schauss_etal12}, and others. On the theoretical side, intriguing physical effects and properties have been predicted for long-range interacting quantum systems, including nonequivalent statistical ensembles and negative response functions \cite{Kastner10,KastnerJSTAT10}, equilibration time scales that diverge with system size \cite{Kastner11,Kastner12,BachelardKastner13}, prethermalization \cite{vdWorm_etal13,GongDuan13}, and others. 

In this article we study the propagation in time and space of various physical quantities, and this is another topic where long-range interactions lead to peculiar behavior. A number of papers devoted to this topic have appeared in the past two years, reporting results on the spreading of correlations, information, or entanglement \cite{EisertvdWormManmanaKastner13,HaukeTagliacozzo13,Hazzard_etal13,Schachenmayer_etal13,Hazzard_etal14,NezhadhaghighiRajabpour14,RajabpourSotiriadis15}. In short-range systems, all these quantities are known to propagate approximately within a soundcone, reminiscent of the lightcone in relativistic theories, with only exponentially small effects outside the cone. This behavior is termed quasilocality and was rigorously proved by Lieb and Robinson for a class of short-range interacting lattice models \cite{LiebRobinson72}. In the presence of long-range interactions this picture is altered significantly: the concept of a group velocity breaks down, and the spreading of correlations, information, or entanglement may speed up dramatically. This, in turn, has a bearing on all kinds of dynamical properties, and one might hope to harness long-range interactions for fast information transmission, improved quantum state transfer, or other applications.

\begin{figure}
{\center
\includegraphics[height=0.25\linewidth]{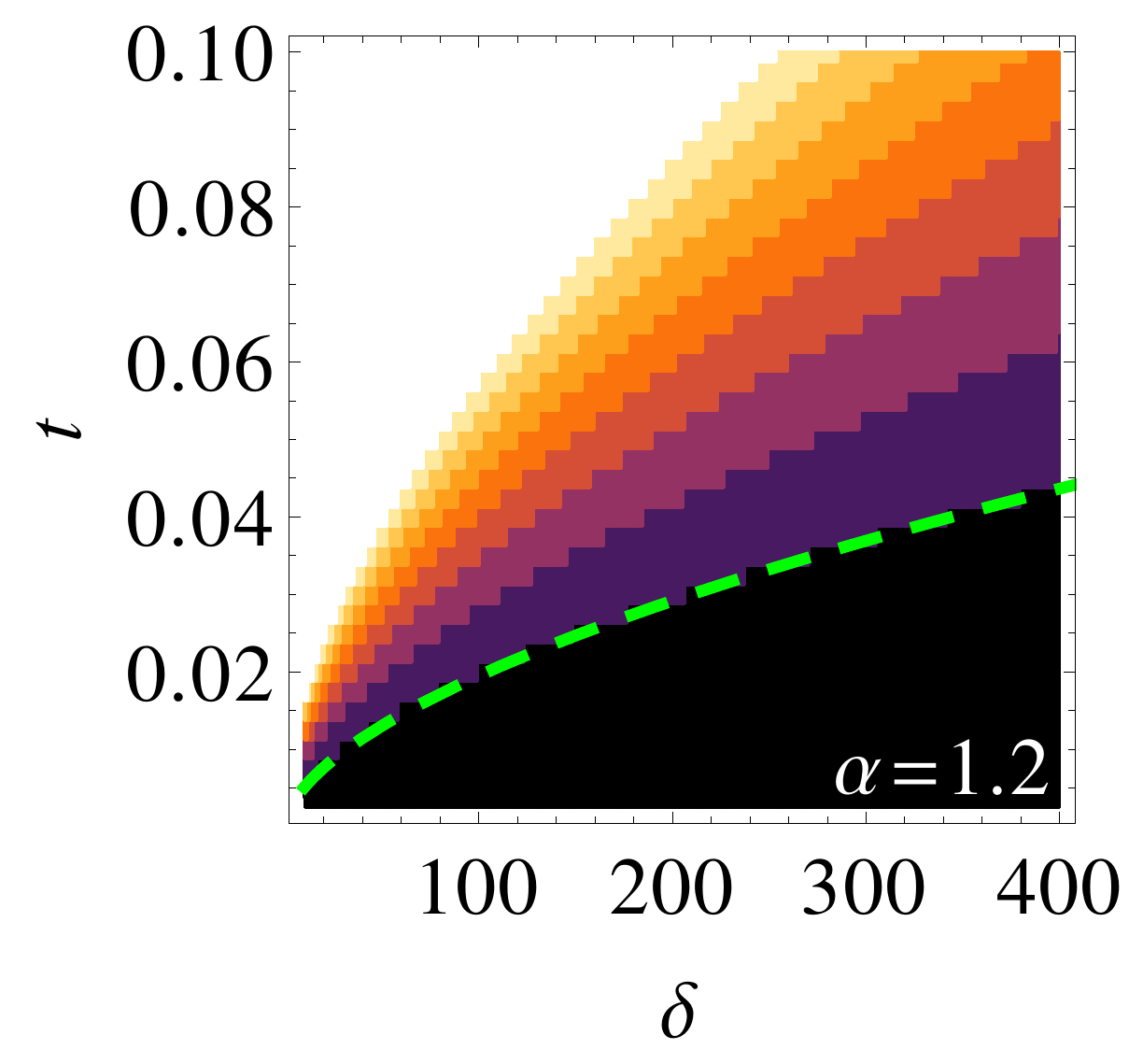}
\includegraphics[trim = 0mm -20mm 0mm -10mm, height=0.25\linewidth]{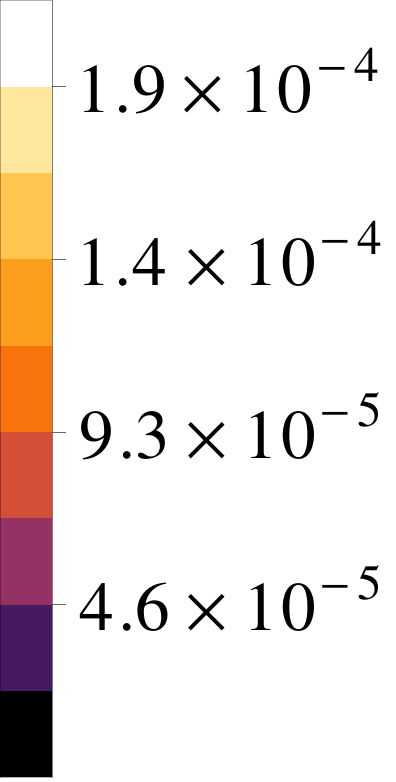}
\hspace{-4mm}
\includegraphics[height=0.25\linewidth]{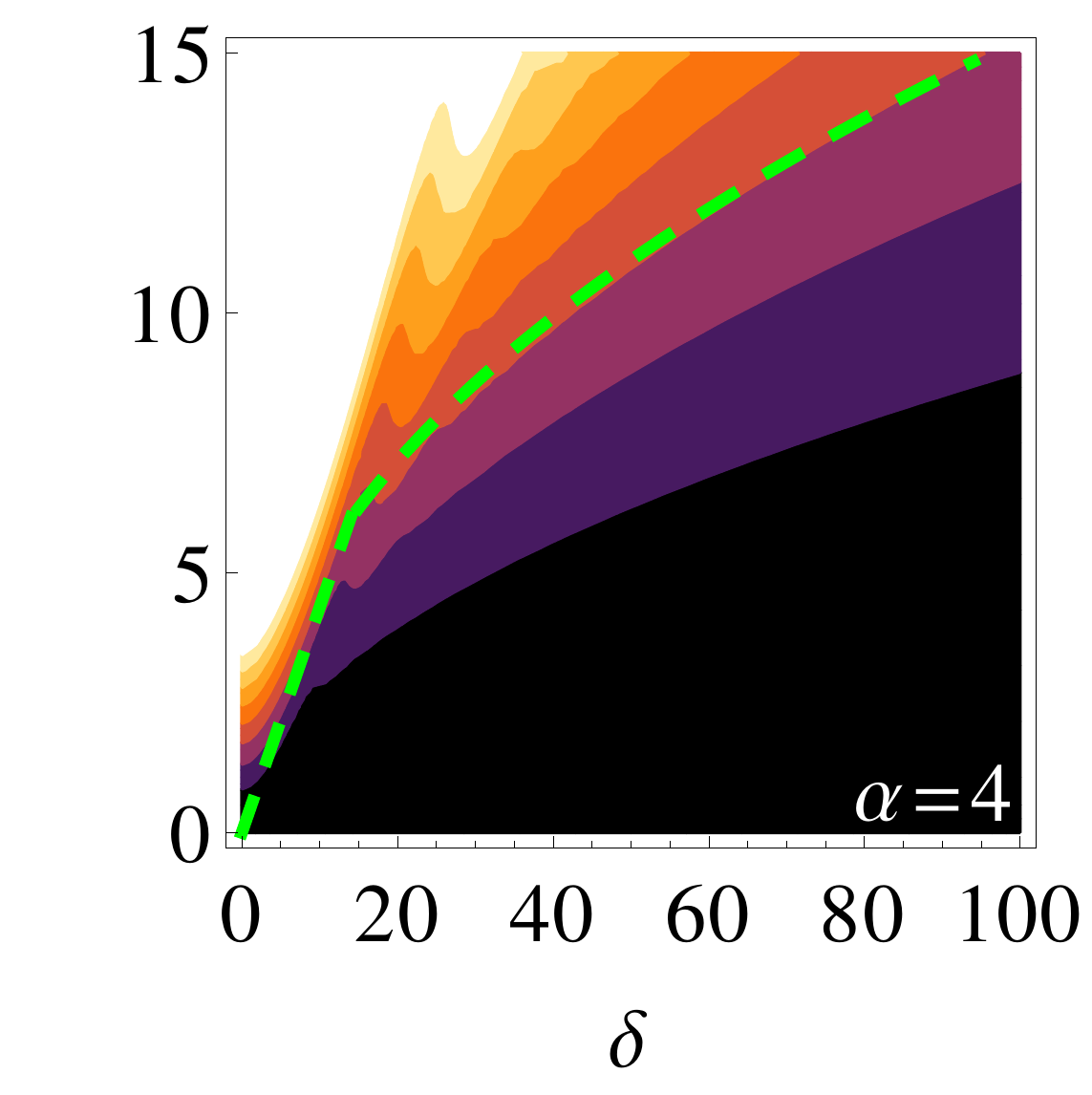}
\includegraphics[trim = 0mm -19mm 0mm -8.5mm, height=0.25\linewidth]{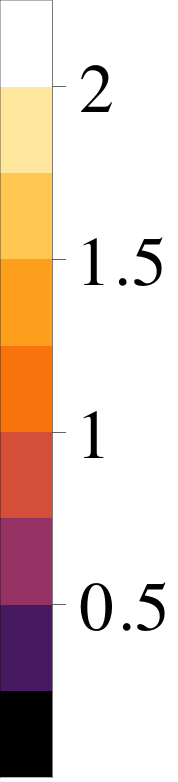}
\hspace{-4mm}
\includegraphics[height=0.25\linewidth]{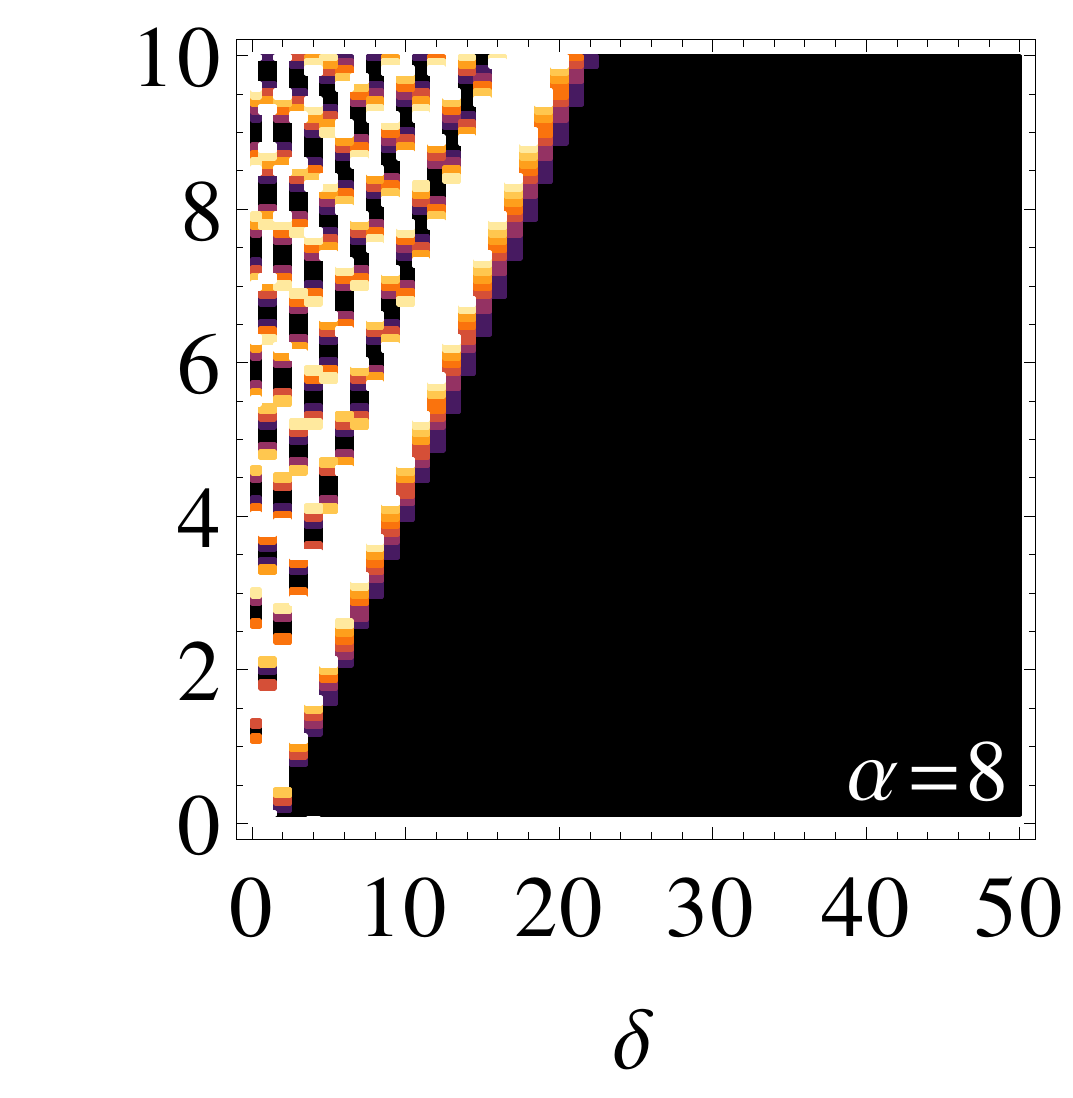}
\includegraphics[trim = 0mm -20mm 0mm -10mm, height=0.25\linewidth]{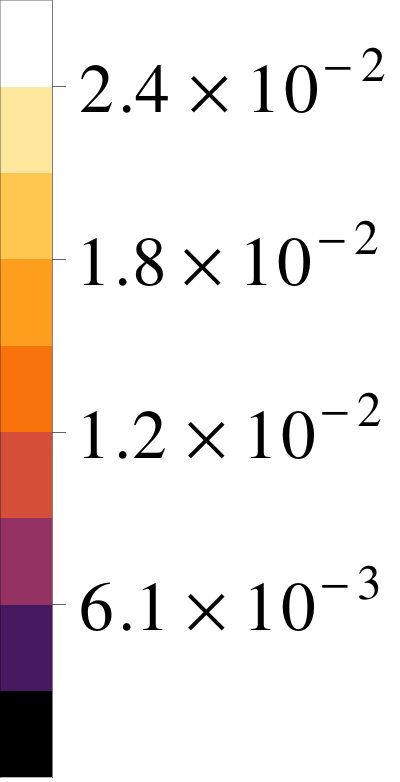}
}
\caption{\label{f:horizon}%
Propagation patterns as a function of distance $\delta=\dist(i,j)$ and time $t$ for different long-range exponents $\alpha$. To highlight the generality of the phenomena we discuss in this article, we use different models and physical quantities as examples. 
Left: For a long-range Ising chain with $\alpha=1.2$, we show the probability to detect a signal sent through a quantum channel from site $0$ to $\delta$ \cite{EisertvdWormManmanaKastner13}. The green line is a guide to the eye and shows a power law $\delta \propto t^{1.7}$. Center: Connected equal-time correlations between lattice sites 0 and $\delta$ in a long-range field theory in one spatial dimension with $\alpha=4$ \cite{RajabpourSotiriadis15}. After an initial cone-like spreading, a cross-over to power law-shaped contours is observed.  The green dashed curve is a guide to the eye. Right: The spreading of entanglement as captured by the mutual information between two lattice sites separated by a distance $\delta$ in the long-range hopping model (\ref{eq:Hopping_Hamiltonian}) with $\alpha=8$, starting from a staggered initial state (see text). Entanglement is sharply confined to the interior of a cone.
}
\end{figure}

Much of our understanding of propagation in long-range systems comes from analytical or numerical studies of model systems, where for example correlations or entanglement between lattice sites $i$ and $j$ are calculated as functions of time $t$ and spatial separation $\dist(i,j)$. Typical examples of such results, similar to some of those in \cite{EisertvdWormManmanaKastner13,HaukeTagliacozzo13,Hazzard_etal13,Schachenmayer_etal13,Hazzard_etal14,NezhadhaghighiRajabpour14,RajabpourSotiriadis15}, are shown in figure~\ref{f:horizon} for a number of different models, physical quantities, and exponents $\alpha$. For larger $\alpha$ (figure~\ref{f:horizon} right), the behavior is reminiscent of the short-range case, with only small effects outside a cone-shaped region. For small $\alpha$ (figure~\ref{f:horizon} left), correlations propagate faster than any finite group velocity would permit, and are mostly confined to a region with power law-shaped boundaries. For intermediate $\alpha$ (figure~\ref{f:horizon} center), a crossover from cone-like to faster-than-cone behavior is observed. While these three regimes seem to be typical and occur in many of the models studied, notable exceptions (some of which will be discussed further below) do occur and lead to a more complicated overall picture.

Besides model calculations, Lieb-Robinson-type bounds have contributed significantly to our understanding of propagation in long-range interacting models. The first result of this kind, 
\begin{equation}\label{e:LRLR}
\left|\left|\left[O_A(t),O_B(0)\right]\right|\right|
\leq C \left|\left| O_A\right|\right| \left|\left| O_B\right|\right|
\frac{\left| A\right|\left| B\right|  (e^{v\left| t\right|}-1)}{[\dist(A,B)+1]^\alpha},
\end{equation}
valid for exponents $\alpha$ larger than the lattice dimension $D$, was reported by Hastings and Koma \cite{HastingsKoma06}. Here, $A,B \subset\Lambda$ are non-overlapping regions of the lattice $\Lambda$, and $O_A(0)$ and $O_B(0)$ are observables supported only on the subspaces of the Hilbert space corresponding to $A$ and $B$, respectively. $\left|\left|\cdot \right|\right|$ denotes the operator norm, and $\dist(A,B)$ is the graph-theoretic distance between $A$ and $B$~\footnote{The graph-theoretic distance is the number of edges along the shortest path connecting the two regions.}. The relevance of the bound (\ref{e:LRLR}) lies in the fact that a number of physically interesting quantities, like equal-time correlation functions, can be related to the operator norm of the commutator on the left-hand side of (\ref{e:LRLR}), so that similar bounds hold also for these physical quantities \cite{BravyiHastingsVerstraete06,NachtergaeleOgataSims06}. For any $\alpha$, a contour plot of the bound \eref{e:LRLR} looks qualitatively like the plot in Fig.~\ref{f:horizon} (left), although with logarithmic contour lines instead of power laws. This implies that, while correct as a bound for all $\alpha>D$, the shape of the propagation front (figure~\ref{f:horizon} center and right) is not correctly reproduced by (\ref{e:LRLR}) for intermediate or large values of $\alpha$. Another bound put forward in \cite{GongFossFeigMichalakisGorshkov14} improves the situation for the case of large $\alpha$, but turns out to be weaker than (\ref{e:LRLR}) for smaller values~\footnote{See \ref{s:AppGong} for a more detailed discussion of the bound in \cite{GongFossFeigMichalakisGorshkov14}.}. Summarizing the situation, the existing Lieb-Robinson-type bounds struggle to reproduce the transition from cone-like to faster-than-cone propagation for intermediate $\alpha$ as in figure~\ref{f:horizon} (center)~\footnote{We could not compare the tightness of the matrix exponential bound with that of the bound in \cite{FossFeigGongClarkGorshkov15}, as several of the constants occurring in that bound were not specified.}. For small $\alpha$, no bounds have been published so far.

In this article we prove general bounds, complemented by model calculations, that help to establish a comprehensive and consistent picture of the various kinds of propagation behavior that occur in long-range interacting lattice models. We extend Lieb-Robinson-type bounds to strong long-range interactions where $\alpha<D$. This is complemented by model calculations showing that, even in the regime $\alpha<D$ of strong long-range interactions, cone-like propagation may be a dominant feature. We also prove that faster-than-cone propagation can occur for all $\alpha<2$ in any spatial dimension, and this answers a question put forward in \cite{Richerme_etal14}. For intermediate exponents $\alpha$, we advocate the use of a Lieb-Robinson-type bound in the form of a matrix exponential, which is tight enough to capture the transition from a cone-like to a faster-than-cone propagation as in figure~\ref{f:horizon} (center), and is also computationally efficient.

\section{\boldmath Lieb-Robinson bounds for \texorpdfstring{$\alpha<D$}{alpha<D}}
\label{s:smallalpha}

For deriving analytical results in the regime $\alpha<D$, an understanding of the time scales of the dynamics turns out to be crucial. The presence of strong long-range interactions is known in many cases to cause a scaling of the relevant time scales with system size \cite{AnRu95,Kastner11,Kastner12,BachelardKastner13,vdWorm_etal13,GongDuan13}. For long-range quantum lattice models the fastest time scale $\mathscr{T}\propto N^{-q}$ was found to shrink like a power law with increasing system size $N$, where $q$ is a positive exponent \cite{BachelardKastner13,vdWorm_etal13}. This observation makes clear why previous attempts to derive a Lieb-Robinson-type bound for $\alpha<D$ failed: in the large-$N$ limit the dynamics becomes increasingly faster, and hence propagation is not bounded by any finite quantity. Considering evolution in rescaled time $\tau=tN^q$ can resolve this problem and allows us to obtain a finite bound in the thermodynamic limit.

\begin{figure}
{\center 
\includegraphics[height=0.3\linewidth]{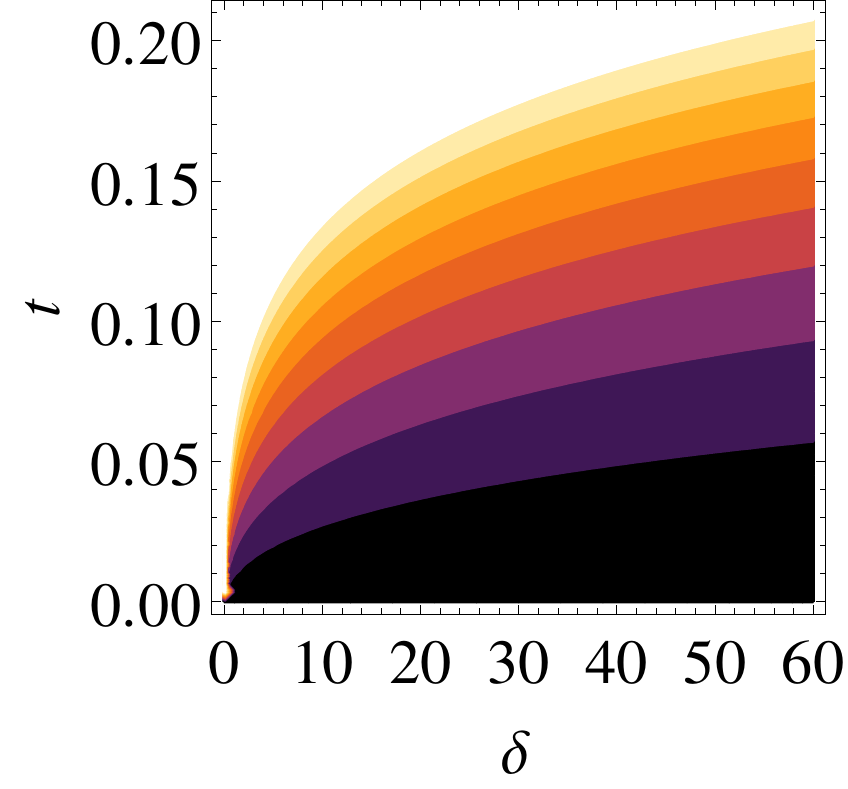}
\hfill
\includegraphics[height=0.3\linewidth]{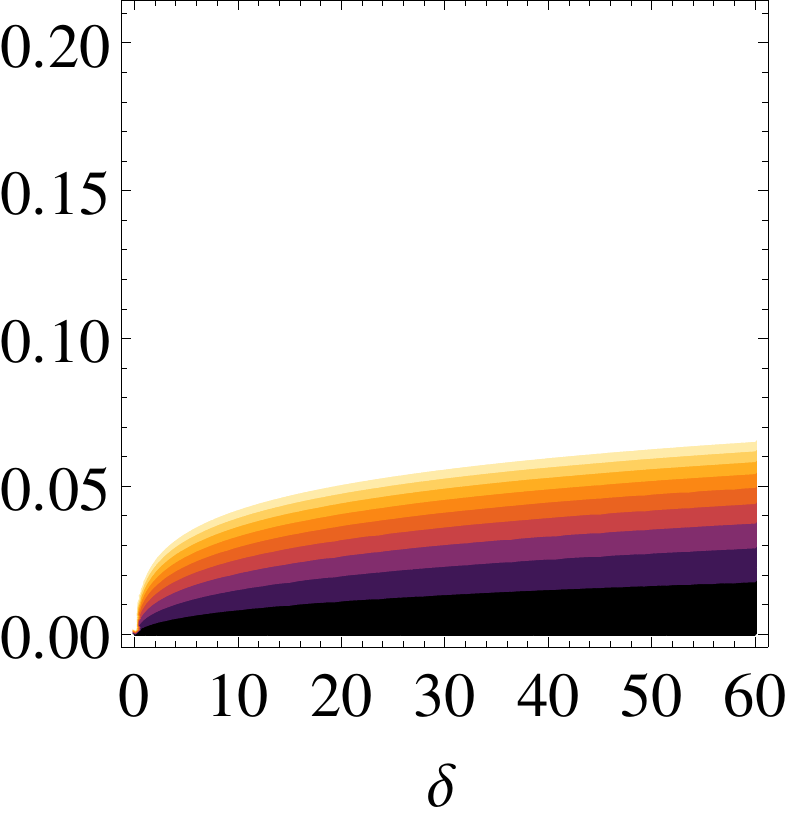}
\hfill
\includegraphics[height=0.3\linewidth]{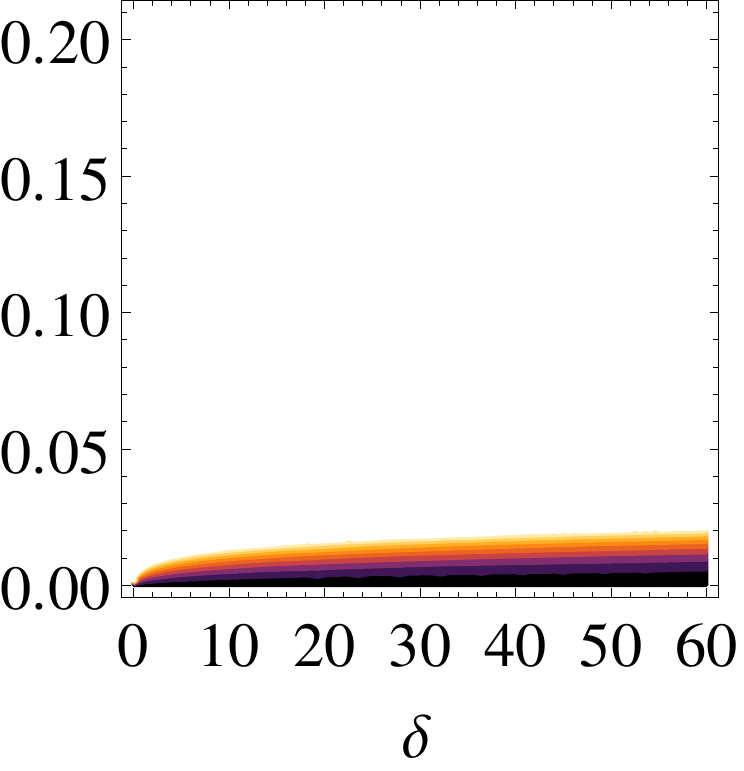}
\hfill
\includegraphics[height=0.3\linewidth]{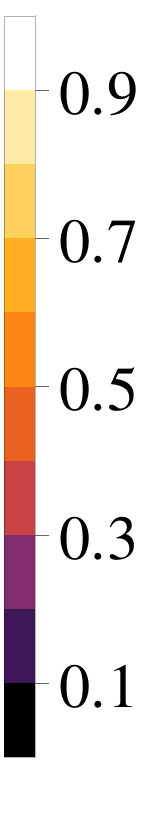}
}
\caption{\label{f:speedup}
Bound (\ref{e:LRBRescaledTime}) in physical (not rescaled) time $t$ for $\alpha=1/2$ and lattice sizes $N=10^2$, $10^3$, and $10^4$ (from left to right), illustrating the speed-up of the propagation with increasing lattice size. For simplicity all constants in (\ref{e:LRBRescaledTime}) are set to unity.}
\end{figure}

On an arbitrary $D$-dimensional lattice $\Lambda$ with $N$ sites we consider the Hilbert space
\begin{equation}\label{e:Hilbert}
\mathscr{H}=\bigotimes_{i=1}^N \mathscr{H}_i
\end{equation}
with finite-dimensional local Hilbert spaces $\mathscr{H}_i$. On $\mathscr{H}$ a  generic Hamiltonian
\begin{equation}\label{e:Hgeneric}
H=\sum_{X\subset\Lambda} h_X
\end{equation}
with $n$-body interactions is defined, with local Hamiltonian terms $h_X$ compactly supported on the finite subsets $X\subset \Lambda$. The Hamiltonian is required to satisfy the following two conditions.
\begin{enumerate}
 \item[(i)] {\em Boundedness},
  \begin{equation}\label{e:Cond1}
\sum_{X\ni i,j} \left|\left| h_X\right|\right| \leq \frac{\lambda}{[1+\dist(i,j)]^\alpha}
 \end{equation}
with a finite constant $\lambda>0$. This condition, also used in \cite{HastingsKoma06}, is a generalization of the definition of power law-decaying interactions, and it reduces to the usual definition in the case of pair interactions, i.e., when $X$ consists only of the two elements $i$ and $j$.
\item[(ii)] {\em Reproducibility},
 \begin{equation}\label{e:Cond2}
\mathscr{N}_\Lambda \sum_{k\in\Lambda} \frac{1}{[1+\dist(i,k)]^\alpha [1+\dist(k,j)]^\alpha}\leq \frac{p}{[1+\dist(i,j)]^\alpha}
 \end{equation}
for finite $p>0$, with
\begin{equation}
 \mathscr{N}_\Lambda=1/\sup_{i\in \Lambda}\sum_{j\in \Lambda\backslash\{i\}} \frac{1}{[1+\dist(i,j)]^\alpha}.
\end{equation}
\end{enumerate}
The lattice-dependent factor $\mathscr{N}_\Lambda$ is the same that is frequently used to make a long-range Hamiltonian extensive \cite{CamDauxRuf09,Kastner11}, but we use it here for a different purpose. Asymptotically for large regular lattices, one finds \cite{Kastner11}
\begin{eqnarray}
\mathscr{N}_\Lambda&\sim & 
\cases{\displaystyle c_1 N^{\alpha/D-1} &\text{for $0\leqslant\alpha<D$},\\
\displaystyle c_2/\ln N &\text{for $\alpha=D$},\\
\displaystyle c_3 &\text{for $\alpha>D$},
}
\end{eqnarray}
with $\alpha$-dependent positive constants $c_1$, $c_2$, and $c_3$. Eq.~(\ref{e:Cond2}) is a modified version of one of the requirements for the proof in \cite{HastingsKoma06}, but due to the modification by the factor $\mathscr{N}_\Lambda$ the condition is satisfied for a larger class of models, including regular $D$-dimensional lattices with power law-decaying interactions with arbitrary positive exponents $\alpha$ \cite{MetivierBachelardKastner14}. For the above described setting we derive in \ref{s:AppRescaledBound} the Lieb-Robinson-type bound
\begin{equation}\label{e:LRBRescaledTime}
\left|\left|\left[O_A(\tau\mathscr{N}_\Lambda),O_B(0)\right]\right|\right|
\leq C \left|\left| O_A\right|\right| \left|\left| O_B\right|\right| 
\frac{\left| A\right|\left| B\right|  (e^{v\left| \tau \right|}-1)}{p[\dist(A,B)+1]^\alpha}
\end{equation}
in rescaled time
\begin{equation}\label{e:tau}
\tau=t/\mathscr{N}_\Lambda.
\end{equation}
This bound reproduces qualitative features of supersonic propagation (as in figure~\ref{f:horizon} left), and also accounts for the system-size dependence of the time scale of propagation for exponents $\alpha<D$. While the bound ensures well-defined dynamics in rescaled time $\tau$ in the thermodynamic limit, it describes a speed-up in physical time $t$ of the propagation with increasing lattice size, as illustrated in figure~\ref{f:speedup}.

\section{\boldmath Matrix exponential bounds for intermediate \texorpdfstring{$\alpha$}{alpha}}
\label{s:MatrixExp}

For long-range models with intermediate exponents, in the range $3<\alpha<6$ or even a bit larger, one observes an interplay of cone-like and supersonic propagation (figure~\ref{f:horizon} center). This is the most relevant regime  for experimental realizations of long-range interactions by means of cold atoms or molecules, but a theoretical description of the shape of the propagation front turns out to be challenging. Existing bounds \cite{GongFossFeigMichalakisGorshkov14} are discussed in \ref{s:AppGong}. Here we report bounds that capture the features of the propagation front as observed in long-range models with intermediate exponents, showing a clear and sharp crossover from cone-like to supersonic propagation. 

As in section~\ref{s:smallalpha}, our setting is a $D$-dimensional lattice $\Lambda$ consisting of $N$ sites and a Hilbert space (\ref{e:Hilbert}) with finite-dimensional local Hilbert spaces. We consider a generic Hamiltonian with pair interactions,
\begin{equation}\label{e:Hpair}
H=\frac{1}{2}\sum_{k\neq l}^N h_{kl},
\end{equation}
where the pair interactions $h_{kl}$ are bounded operators supported on lattice sites $k$ and $l$ only. As observables $O_A$ and $O_B$ we consider bounded operators that are supported on single sites $A=\{i\}$ and $B=\{j\}$. In this setting, we prove in \ref{s:AppMatrixExp} a bound in the form of an $N\times N$ matrix exponential,
\begin{equation}\label{e:MatrixExp}
 \left|\left|\left[O_i(t),O_j(0)\right]\right|\right| \leq 2\left|\left| O_i\right|\right| \left|\left| O_j\right|\right| \left(\exp\left[2 \kappa J \left|t\right| \right]_{i,j} - \delta_{i,j}\right),
\end{equation}
where $J$ is the interaction matrix with elements
\begin{equation}
J_{k,l}=\|h_{kl}\|
\end{equation}
and $\kappa= \sup_{n\in \Lambda} \sum_k J_{n,k}$. In one-dimensional homogeneous lattice models the interaction matrix $J$ is of Toeplitz type and thus (\ref{e:MatrixExp}) can be evaluated in $\mathscr{O}(N^2)$ time using the Levinson algorithm \cite{Bareiss69}. For translationally invariant one-dimensional systems, $J$ is a circulant matrix, which permits an analytical solution of (\ref{e:MatrixExp}) by means of Fourier transformation.

\begin{figure}
{\center
\includegraphics[height=0.285\linewidth]{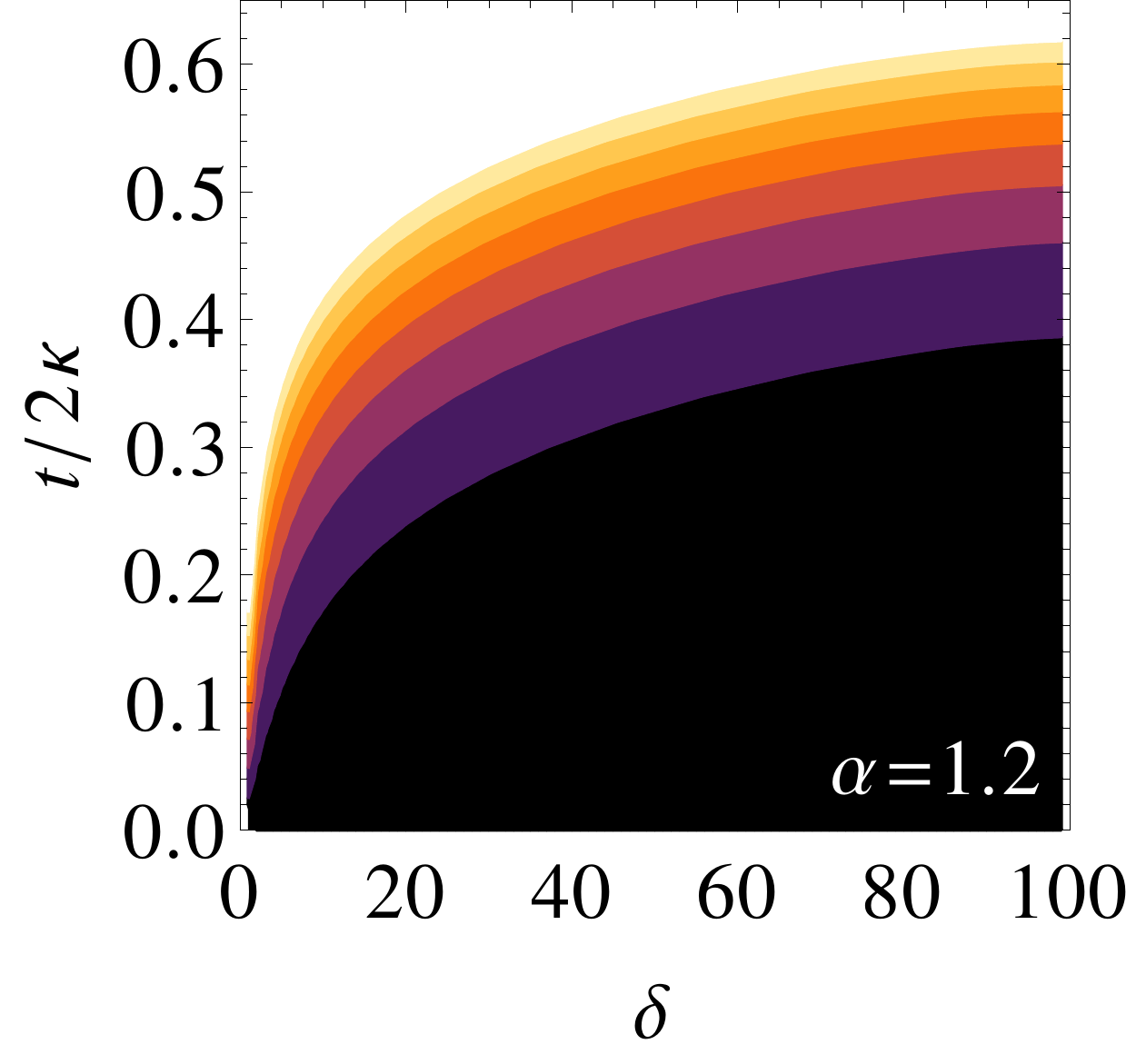}
\hfill
\includegraphics[height=0.285\linewidth]{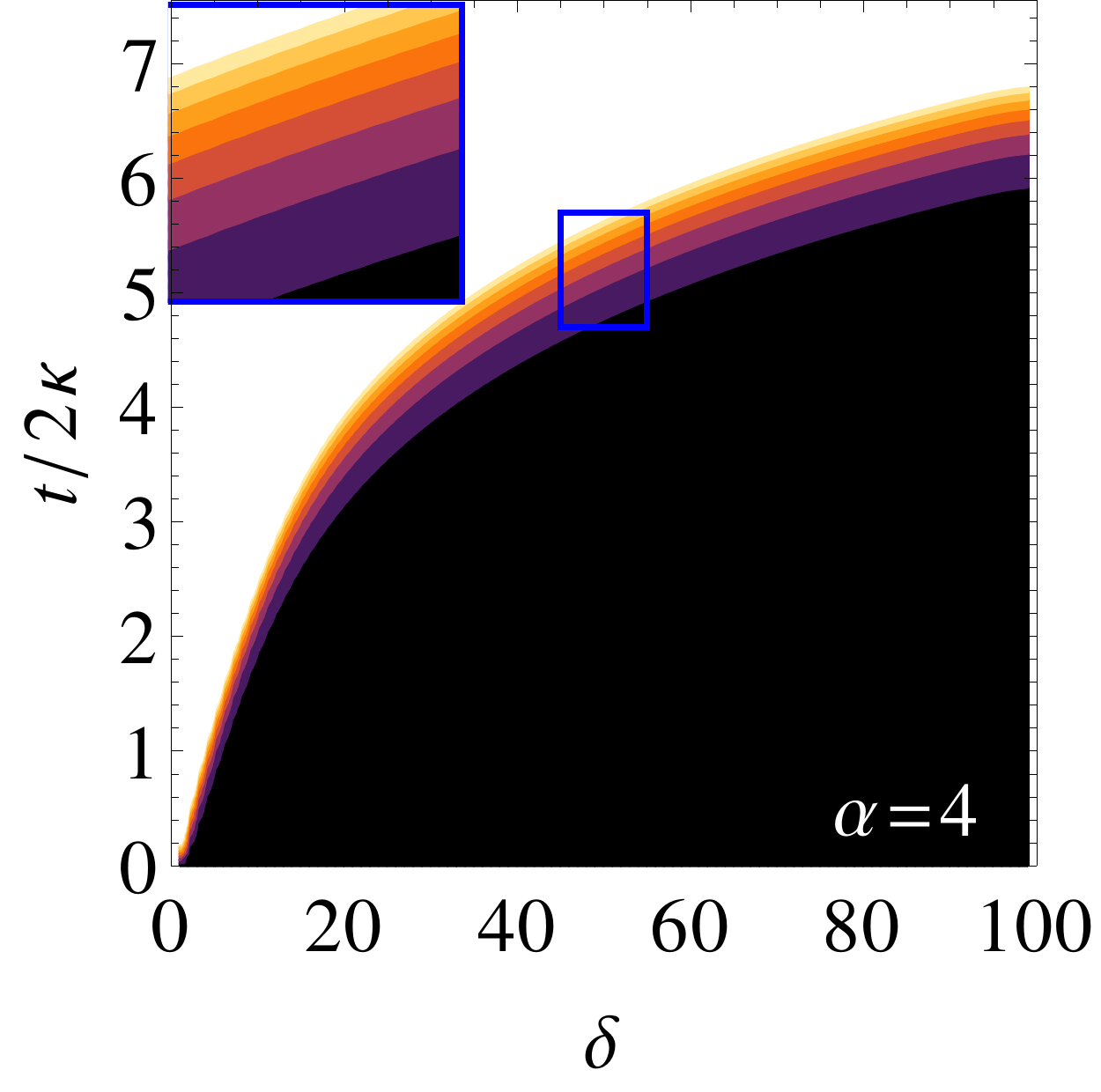}
\hfill
\includegraphics[height=0.285\linewidth]{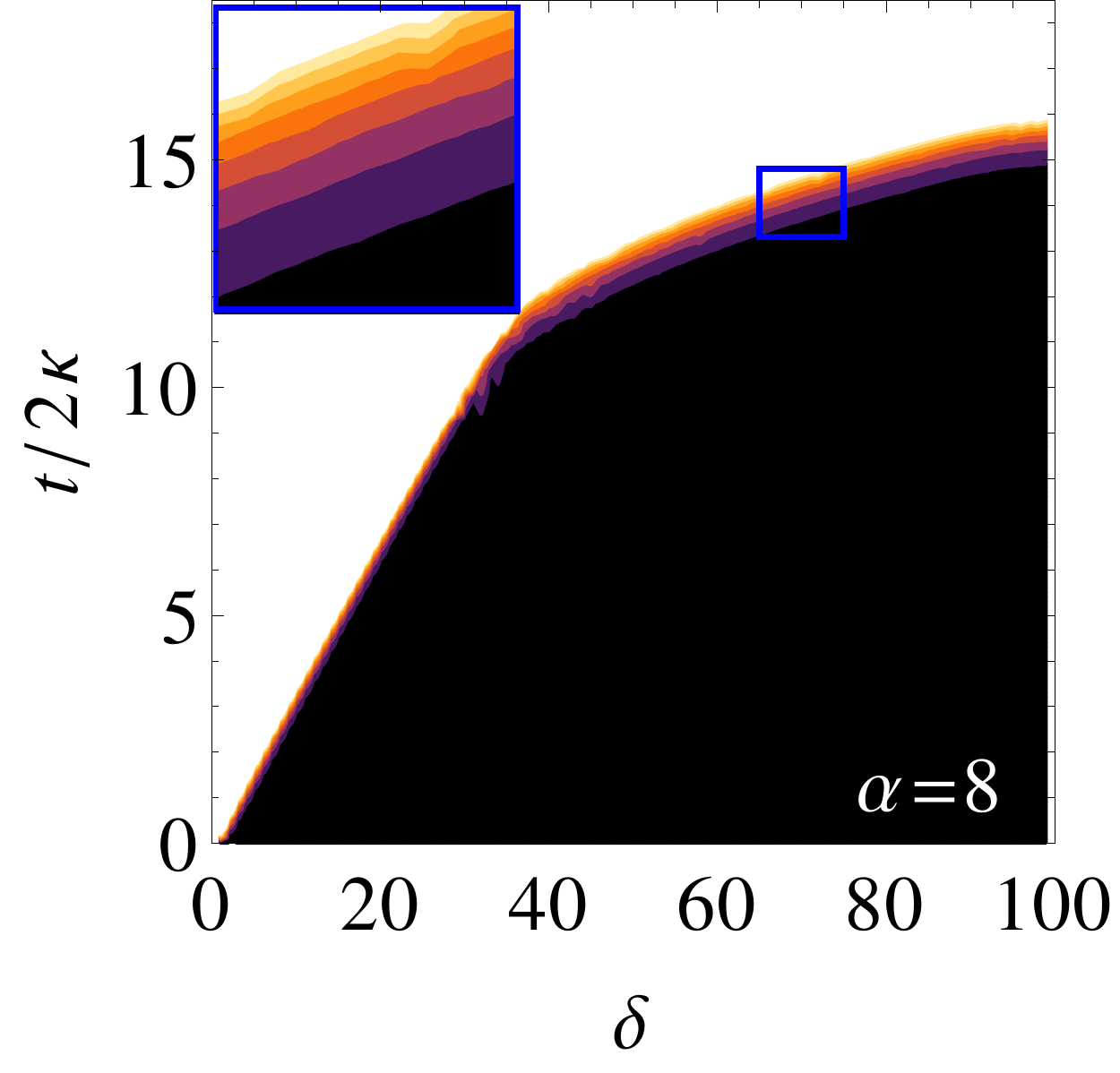}
\hfill
\includegraphics[height=0.285\linewidth]{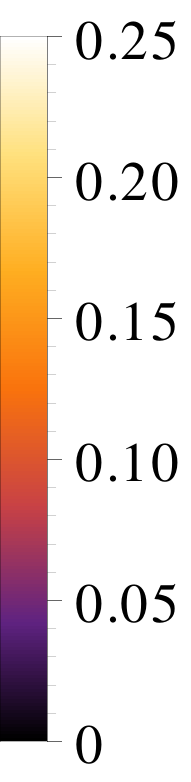}
}
\caption{\label{f:MExp}
Spacetime plots of the matrix exponential bound (\ref{e:MatrixExp}) for several values of $\alpha$ in a one dimensional system with $L=201$ lattice sites and periodic boundary conditions. Left: for $\alpha=1.2$ the bound recovers a propagation front with a shape similar to the one of the Ising model in figure~\ref{f:horizon} (left). Center: for intermediate $\alpha=4$ a transition from soundcone to supersonic dynamics is being heralded. Right: the two regimes of soundcone-like and supersonic dynamics are fully exposed for $\alpha=8$.}
\end{figure}

The bound (\ref{e:MatrixExp}) is tighter than the bounds in \cite{HastingsKoma06,NachtergaeleOgataSims06,GongFossFeigMichalakisGorshkov14}, and the crossover from cone-like to supersonic propagation is nicely captured (see figure~\ref{f:MExp}). Due to its form as a matrix exponential, the bound is less explicit than others, and asymptotic properties are not easily read off. But since the calculation of a matrix exponential scales polynomially in the matrix dimension $N$ (like $\mathscr{O}(N^3)$ or even faster \cite{MolerVanLoan03}) the bound can easily be evaluated for large lattices up to $\mathscr{O}(10^4)$ on a desktop computer. This is orders of magnitude larger than the sizes that can be treated by exact diagonalization, and covers the system sizes that can be reached for example with state-of-the-art ion trap based quantum simulators of spin systems \cite{Britton_etal12}. Different from other bounds of Lieb-Robinson-type, our matrix exponential bound is computed for the exact type of interaction matrix realized in a specific experimental setup. This improves the sharpness of the bound, and can make it a useful tool for investigating all kinds of propagation phenomena in lattice models of intermediate system size.

\section{\boldmath Long-range hopping for small \texorpdfstring{$\alpha$}{alpha}}

The bounds discussed in sections~\ref{s:smallalpha} and \ref{s:MatrixExp} are valid for arbitrary initial states, and therefore it may well happen that propagation for a given model and some, or even most, initial states is significantly slower than what the bound suggests. Indeed, linear (cone-like) propagation was observed in model calculations even for moderately large exponents like $\alpha=3$ \cite{Hazzard_etal13,HaukeTagliacozzo13,Schachenmayer_etal13,EisertvdWormManmanaKastner13,Hazzard_etal14}. But, as we show in the following, such cone-like propagation can, for suitably chosen initial states, even persist into the strongly long-range regime $0<\alpha<D$. In this and the next section we analyze free fermions on a one-dimensional lattice with long-range hopping, which is arguably the simplest model to illustrate cone-like propagation in long-range models and to explain the observation on the basis of dispersion relations and density of states. While strictly speaking such a long-range hopping model does not meet the conditions under which Lieb-Robinson bounds have been proved, it proves helpful for understanding the conditions under which cone-like propagation may or may not be observed in other long-range interacting models.

\subsection{Long-range hopping model}

Consider a free fermionic hopping model in one dimension with periodic boundary conditions,
\begin{equation}\label{eq:Hopping_Hamiltonian}
H= - \frac{1}{2}\sum^{N}_{j =1} \sum^{N-1}_{l=1}  \dist^{-\alpha}_{l} \left( c^{\dagger}_{j} c_{j+l} + c^{\dagger}_{j+l} c_j \right),
\end{equation}
where $c^{\dagger}_j$, $c_j$ are fermionic creation and annihilation operators at site $j$. We choose long-range hopping rates $\propto \dist^{-\alpha}_{l}$, where
\begin{eqnarray}
\dist_{l} &=& 
\cases{\displaystyle l &\text{if $l\leq N/2$},\\
\displaystyle N-l & \text{if $l>N/2$},}
\end{eqnarray}
is the shortest distance between two sites on a chain with periodic boundary conditions.
A Fourier transformation brings the Hamiltonian into diagonal form
\begin{equation}
H=\sum_k\epsilon(k) a_k^\dagger a_k
\end{equation}
with
\begin{equation}\label{eq:Creation}
 c_j =
 \frac{1}{\sqrt{N}}\sum_{k} \ee^{ \ii kj} a_{k}.
\end{equation}
and dispersion relation
\begin{equation}\label{eq:DispersionRelation}
\epsilon(k)=
- \sum^{N-1}_{l=1} \frac{\cos\left( k l \right)}{d_l^{\alpha}},
\end{equation}
where $k=  2 \pi m/N$ with $m=1,\dots,N$.

\subsection{Propagation from staggered initial state}

\begin{figure}
{\center
 \includegraphics[width=0.46\linewidth]{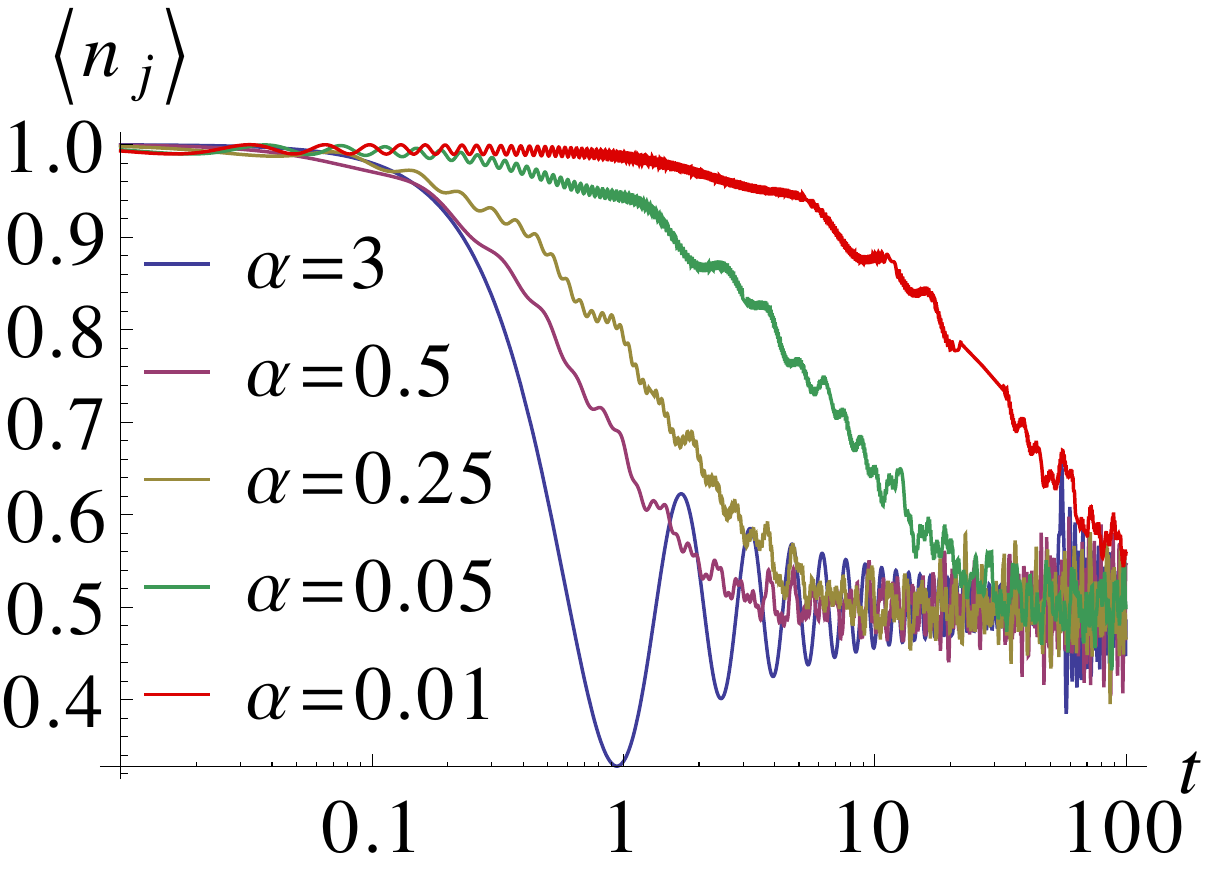}
 \hfill
 \includegraphics[width=0.46\linewidth]{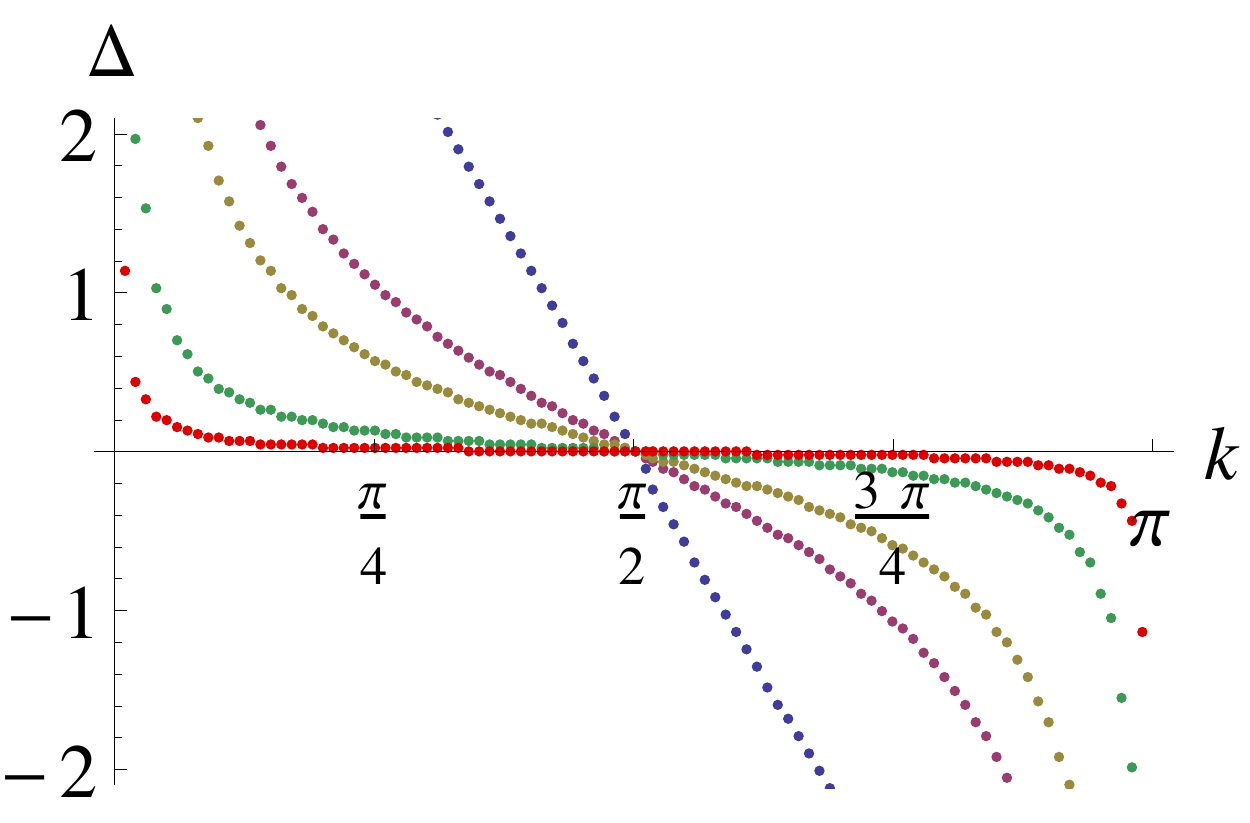}
 }
 \caption{Left: Time dependence of the occupation number of site $j$ for different $\alpha$, starting from a staggered initial state.
 Right: $\Delta$ as a function of $k$.  The system size is $N=200$ in both plots.
}
 \label{fig:NumberOperator}
\end{figure}

We choose a staggered initial state $|1010\dots\rangle$ in position space, i.e., initially every second site is occupied.  For simplicity of notation we assume the number $N$ of lattice sites to be even. A straightforward calculation, similar to that in \cite{CramerEisertScholl08}, yields
\begin{equation}
 \langle n_j(t) \rangle = \frac{1}{2} - \frac{(-1)^{j}}{2N} \sum^N_{n=1} \cos \left[ t \Delta(k) \right]
 \label{eq:NumberOperator} 
\end{equation}
for the time-dependence of the occupation number at lattice site $j$, where
\begin{equation}\label{e:Delta}
\Delta(k):=\epsilon(k+\pi) - \epsilon(k) =  2\sum^{N/2 }_{l=1}\frac{\cos\left[k (2l-1) \right]}{ d_{2l-1}^{\alpha}}
\end{equation}
and $k=2\pi m/N$ with $m=1,\dots,N$. In figure~\ref{fig:NumberOperator} (left) the time evolution of the occupation number $\langle n_j(t) \rangle$ is plotted for different values of $\alpha$, showing that the time it takes to relax to the equilibrium value of $1/2$ increases dramatically for small $\alpha$ (note the logarithmic timescale). This may seem counterintuitive, as a longer interaction range may naively be expected to lead to faster propagation. The effect can be understood from figure~\ref{fig:NumberOperator} (right), showing the spectrum of the frequencies $\Delta$ in the cosine terms of Eq.~(\ref{eq:NumberOperator}). As $\alpha$ decreases, the majority of these frequencies lie within a small window around zero, implying very slow dephasing of the cosine terms.

\begin{figure}[t]
 \centering
 \includegraphics[width=0.28\linewidth]{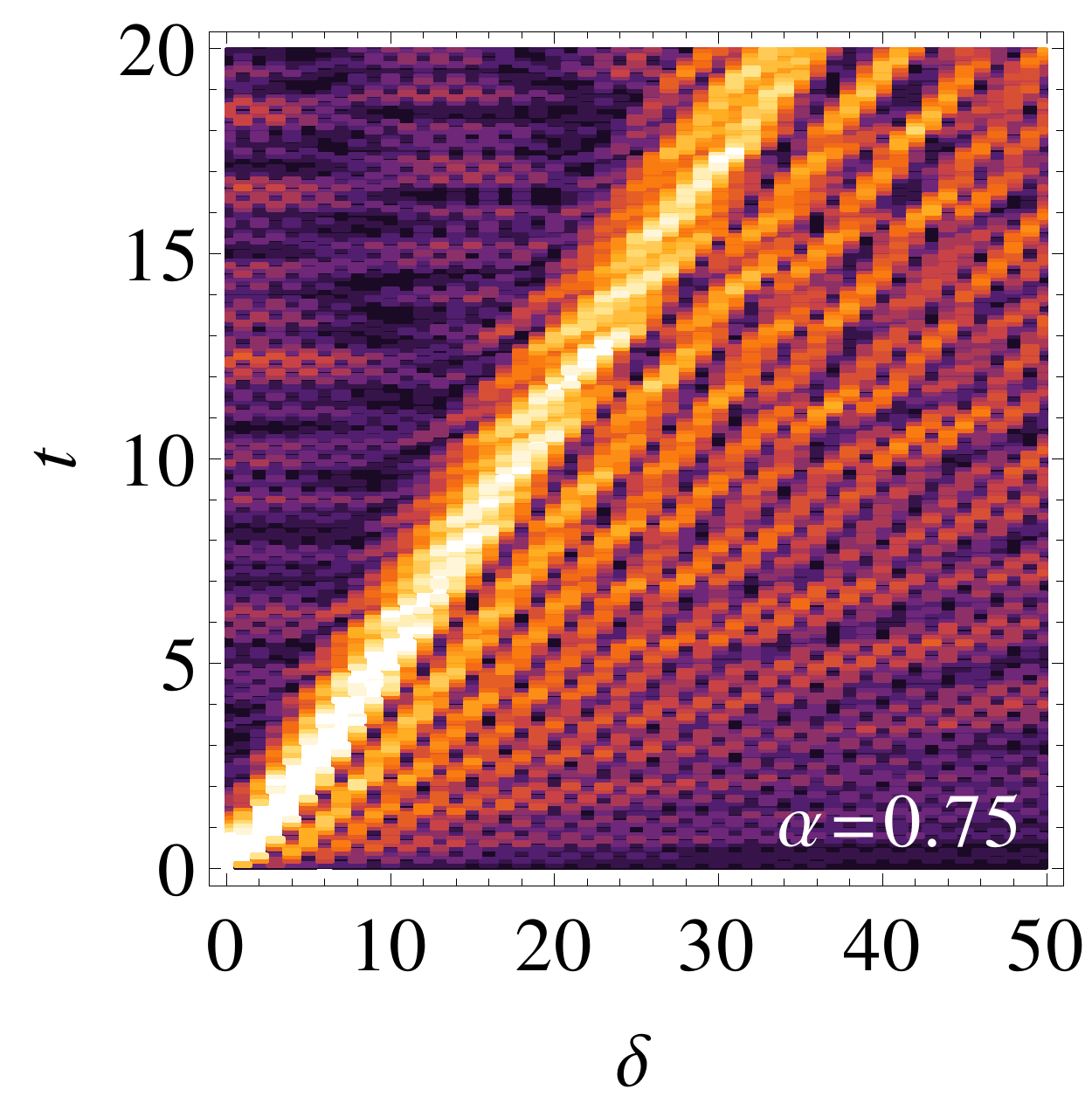} 
 \hfil
 \includegraphics[width=0.28\linewidth]{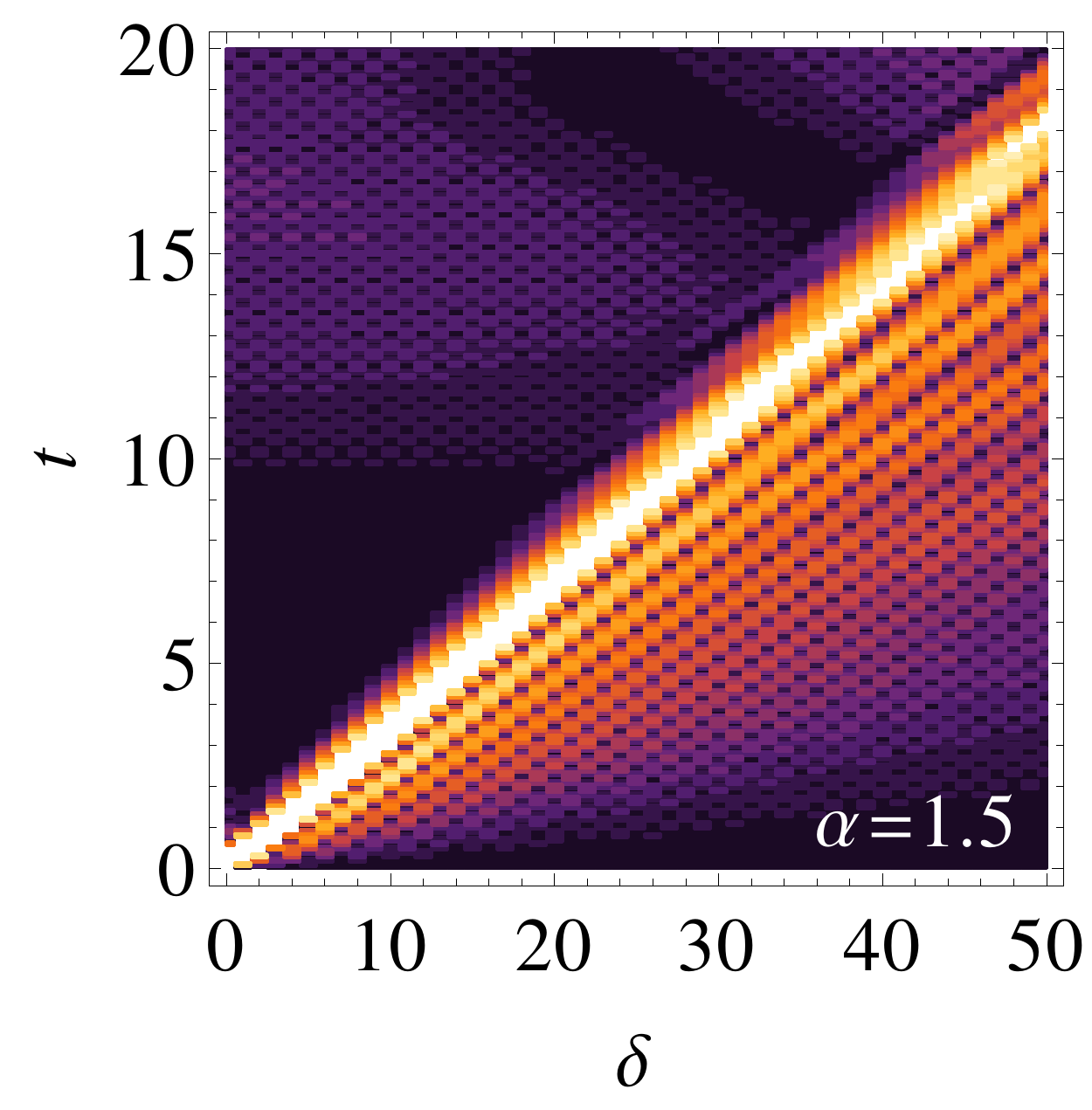} 
 \hfil
 \includegraphics[width=0.28\linewidth]{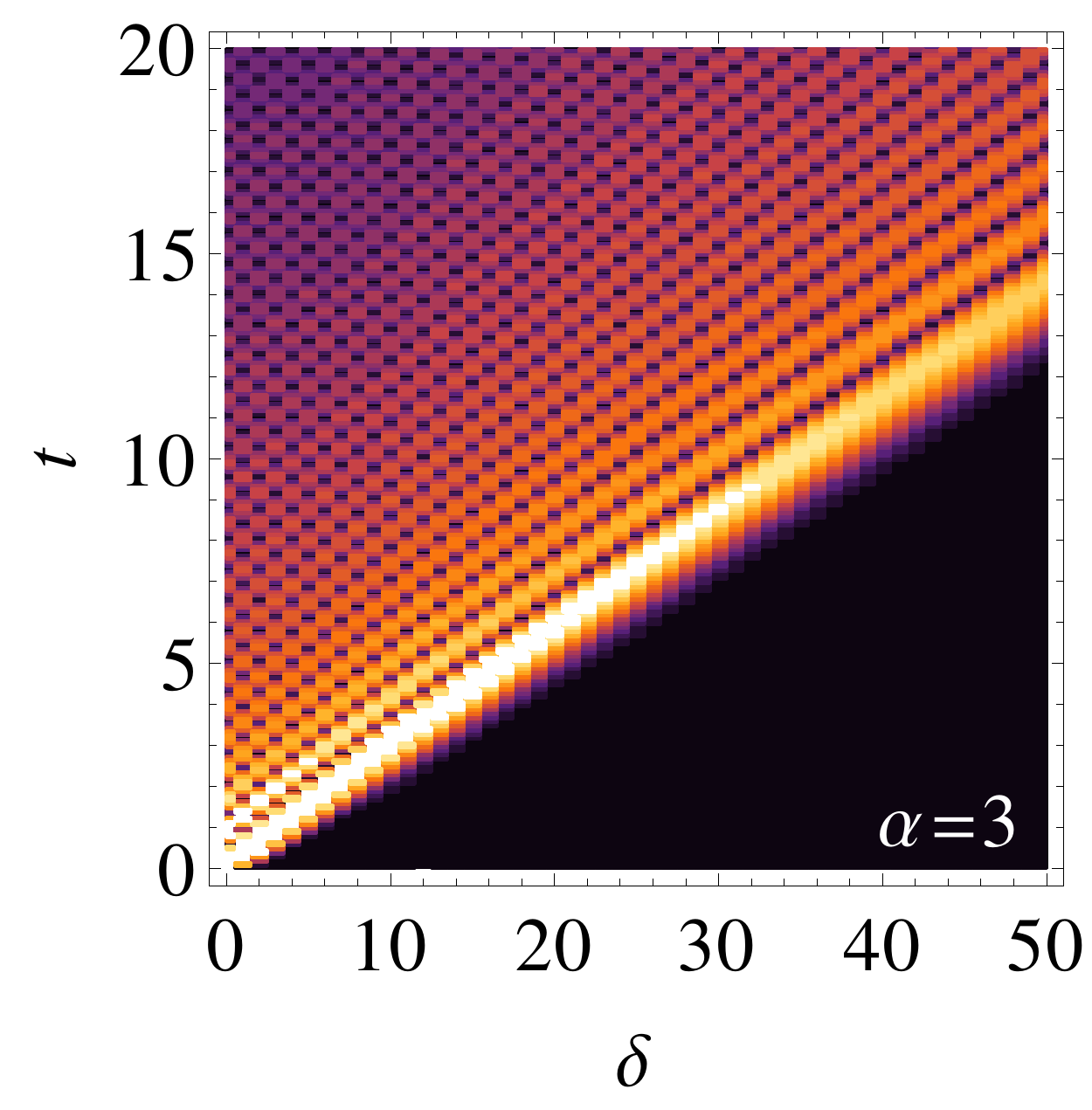} 
 \hfil
 \includegraphics[trim = 0mm -15mm 0mm -20mm, width=0.075\linewidth]{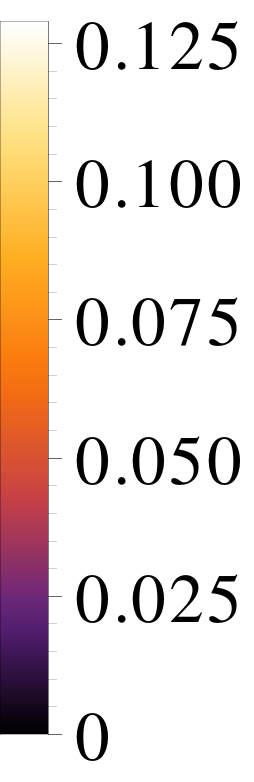} 
 \caption{Contour plots in the $(\delta,t)$-plane, showing correlations (\ref{eq:Staggered_Correlations}) between sites 0 and $\delta$ in the fermionic long-range hopping model for $N=200$ lattice sites and various values of $\alpha$, starting from a staggered initial state. 
 }
 \label{fig:kDependantCone}
\end{figure}

\begin{figure}[b]
{\center
 \includegraphics[width=0.46\linewidth]{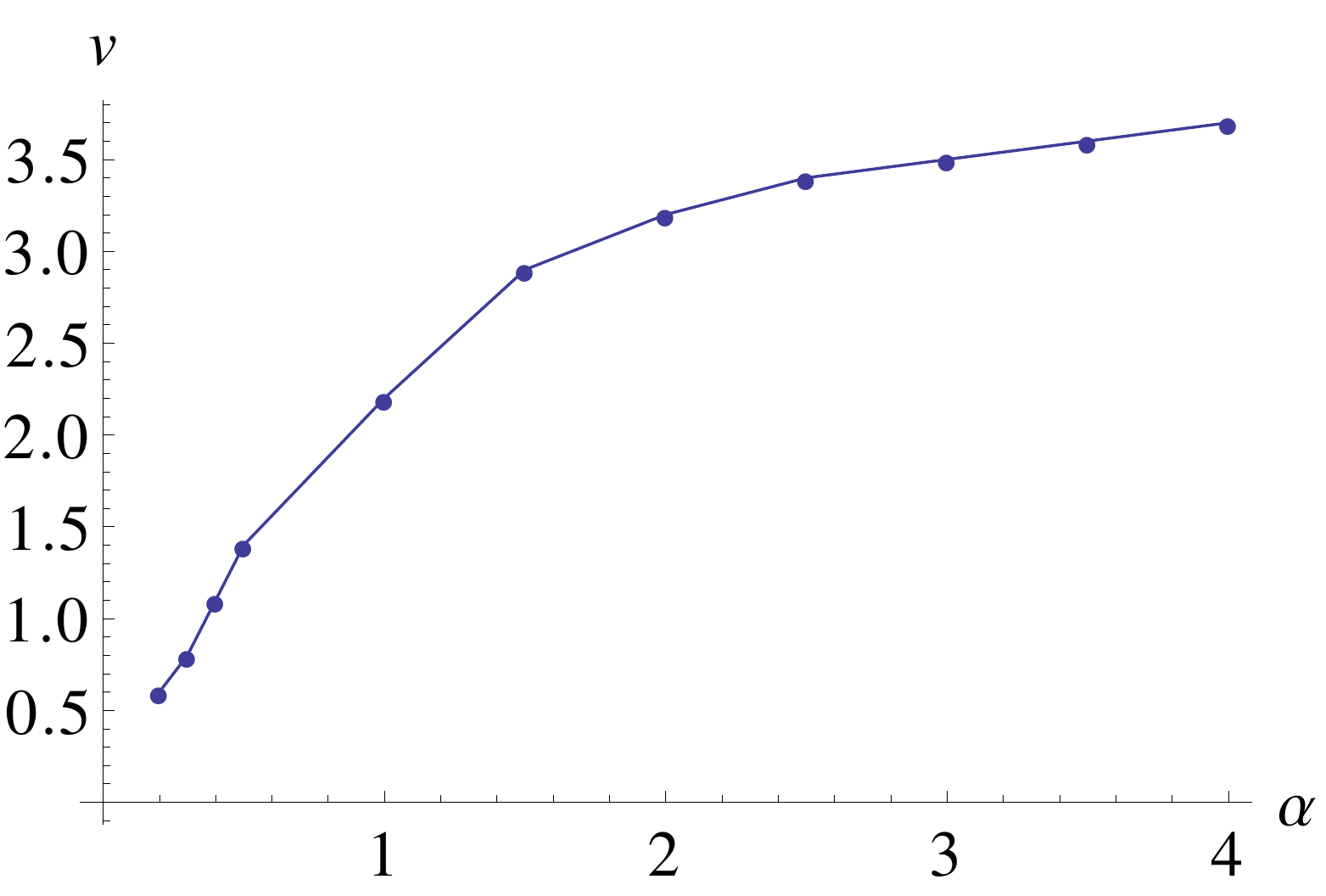}
\hfill
 \includegraphics[width=0.46\linewidth]{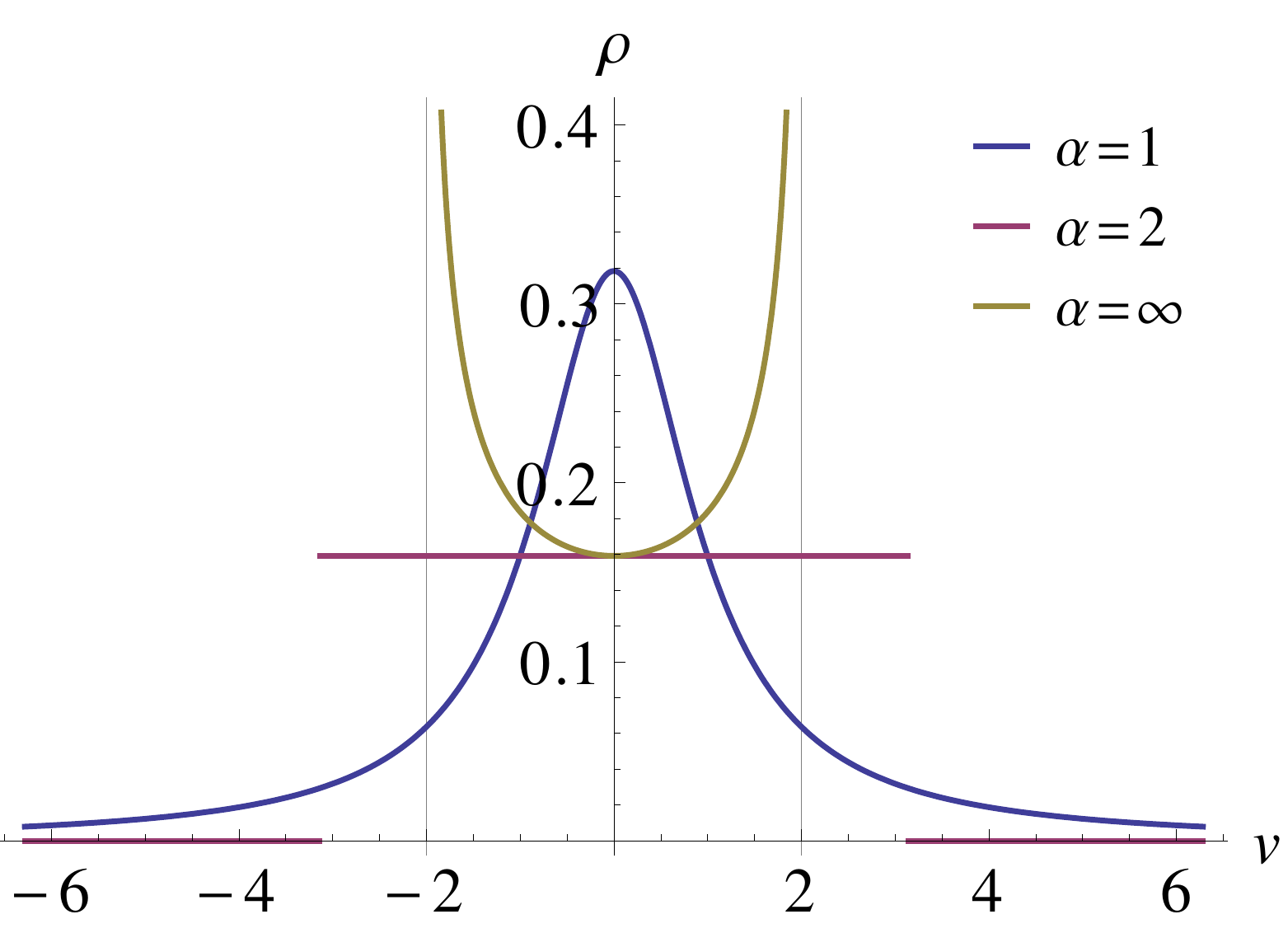}
 }
 \caption{Left: Dominant velocity of propagation, as read off from the inverse slope of the striking cones in figure~\ref{fig:kDependantCone}, plotted as a function of the exponent $\alpha$. Right: Density of states (\ref{eq:DOS01}) for $\alpha=1$, 2 and $\infty$.
 }
 \label{fig:DOS}
\end{figure}

A more refined picture of the propagation behavior can be obtained by studying the spreading of correlations. Starting again from a staggered initial state, a straightforward calculation similar to that in \cite{Cramer2008,CramerEisertScholl08}, and similar to the one leading to (\ref{eq:NumberOperator}), yields
\begin{equation}\label{eq:Staggered_Correlations}
 \langle c^{\dagger}_{j+\delta}(t) c_{j}(t)  \rangle =
\frac{1}{2}\delta_{\delta,0} 
- \frac{(-1)^{j+\delta}}{2N} \sum_{k}
 \ee^{\ii t\left[\epsilon(k+\pi) - \epsilon(k) \right]} \ee^{-\ii k\delta} .
\end{equation}
Figure~\ref{fig:kDependantCone} (bottom) shows contour plots in the $(\delta,t)$-plane of the absolute values of the correlations (\ref{eq:Staggered_Correlations}) for different values of $\alpha$. For all $\alpha$ shown, a cone-like propagation front is clearly visible, even in the case of $\alpha=3/4<D$. Two properties of the cone can be observed to change upon variation of the exponent $\alpha$: (i) The boundary of the cone is rather sharp for larger $\alpha$ (like $\alpha=3$), whereas correlations ``leak'' into the exterior of the cone for smaller $\alpha$ (like $\alpha=3/4$). (ii) The velocity of propagation, corresponding to the inverse slope of the cone, decreases with decreasing $\alpha$ [see figure~\ref{fig:DOS} (left)], confirming the counterintuitive observations of figure~\ref{fig:NumberOperator} (left). We will argue in section~\ref{s:slowdown} that some of these features can be understood on the basis of the dispersion relation (\ref{eq:DispersionRelation}) and the density of states of the long-range hopping model.

\subsection{Dispersion and group velocity}
\label{s:groupvelocity}

\begin{figure}
\centering
 \includegraphics[width=\linewidth]{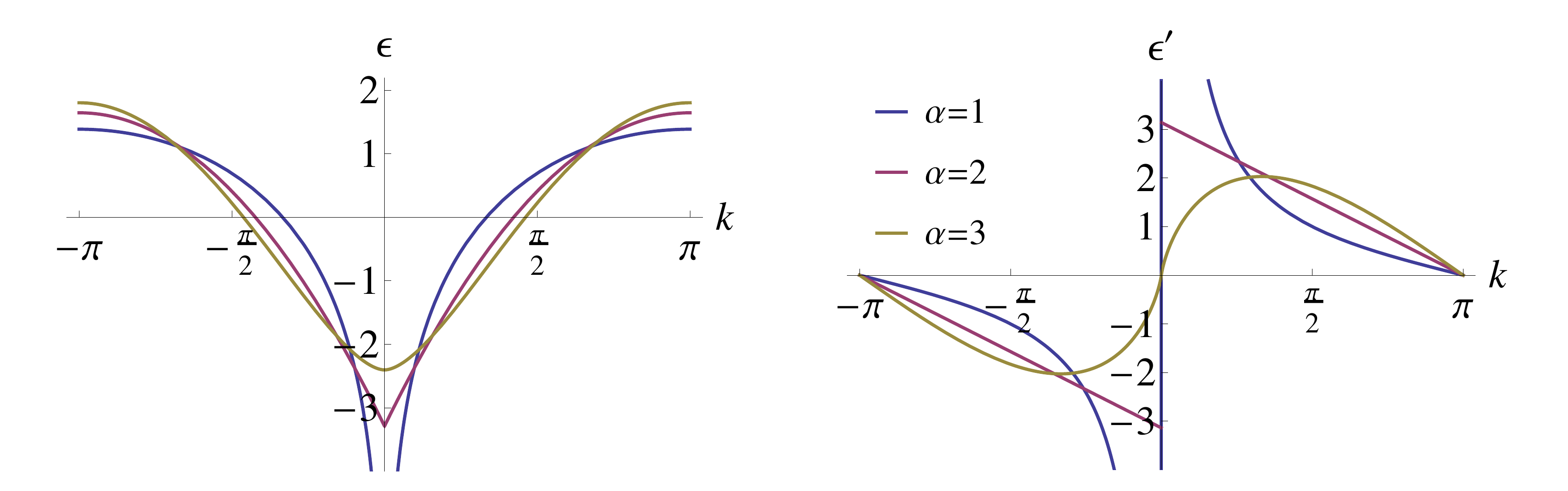}
 \caption{Dispersion relation (\ref{eq:DispersionRelationBigSystem}) (left) and its derivative $\epsilon^\prime(k)$ (right) for the long-range fermionic hopping model (\ref{eq:Hopping_Hamiltonian}) with exponents $\alpha=1$, $2$, and $3$.
}
 \label{fig:DispersionRelation}
\end{figure}

In the limit of large system size the dispersion relation takes the form
\begin{equation}\label{eq:DispersionRelationBigSystem}
 \epsilon(k)=
 -  \left[ \mathrm{Li}_{\alpha}\left( \ee^{\ii k} \right) +  \mathrm{Li}_{\alpha}\left( \ee^{-\ii k} \right) \right] 
\end{equation}
where $\mathrm{Li}_{\alpha}$ is the polylogarithm \cite{NIST}, and this function is plotted in figure~\ref{fig:DispersionRelation} (left). For $\alpha=3$ the dispersion $\epsilon$ is a smooth function of $k$, while it shows a cusp at $k=0$ for $\alpha=2$, and a divergence at $k=0$ for $\alpha=1$. Correspondingly, the derivative $\epsilon'(k)$ as shown in figure~\ref{fig:DispersionRelation} (right) is discontinuous at $k=0$ for $\alpha=2$, and diverges at $k=0$ for $\alpha=1$. More generally we can analyze $\epsilon'$ in the vicinity of $k=0$ by considering the difference quotient between the zeroth and the first mode,
\begin{eqnarray}
 \left|\frac{\epsilon(2\pi/N)-\epsilon(0)}{2\pi(1-0)/N} \right|
 =
 \frac{N}{2\pi} \sum^{N-1}_{l=1} \frac{\left| \cos\left( 2\pi l/N \right)-1 \right|}{d_l^{\alpha}} 
 \\
 \geq
 \frac{N}{2\pi} \sum^{N-1}_{l=1} \frac{\left(2\pi l/N \right)^2}{d_l^{\alpha}} 
 =
 \frac{4\pi}{N} \sum^{N/2}_{l=1} l^{2-\alpha}.
\end{eqnarray}
In the large-$N$ limit we approximate the sum by an integral,
\begin{equation}
 \frac{4\pi}{N} \int^{N/2}_{1} l^{2-\alpha} \dd l = \frac{2\pi}{N(3-\alpha)} \left[ (N/2)^{3-\alpha} - 1 \right]
 \sim N^{2-\alpha}.
\end{equation}
This implies that, for $\alpha<2$, the derivative $\epsilon'$ diverges at $k=0$ in the limit of infinite system size. Interpreting $\epsilon'(0)$ as a group velocity, we infer that we have a finite group velocity only for $\alpha>2$, whereas the concept of a group velocity breaks down for $\alpha<2$~\footnote{The same conclusions about dispersion relations and group velocities also hold for long-range interacting $XX$ and $XXZ$ spin models when restricting the dynamics to the single magnon sector, as the dispersion relations of these models are essentially identical to (\ref{eq:DispersionRelation}).}. This finding can help us to understand figure~\ref{fig:kDependantCone}: For $\alpha>2$ a finite group velocity restricts the propagation to the interior of a cone, which makes this cone appear rather sharp. For $\alpha<2$, although a cone is still visible, larger (and, in fact, arbitrarily large) propagation velocities may occur and are responsible for the ``leaking" of correlations outside the cone.

The threshold value $\alpha=2$ for supersonic propagation (i.e., propagation not bounded by any finite group velocity) is also found in a different context, by very different methods. In \cite{EisertvdWormManmanaKastner13} it was proved that information can be transferred supersonically through a quantum channel with finite local Hilbert space dimension for any $\alpha<D+1$, while no such proof exists for $\alpha>D+1$~\footnote{For models with infinite dimensional local Hilbert spaces $\mathscr{H}_i$, supersonic propagation can occur also in models with nearest-neighbor interactions, although this appears to happen only under rather specific circumstances \cite{EisertGross09}.}, but this result requires the measuring of observables supported on semi-infinite sublattices, which is not the most physical scenario. In \ref{s:AppChannel} we use techniques similar to those in \cite{EisertvdWormManmanaKastner13}, but apply them to a model with translationally invariant interactions, to prove that supersonic transmission through a quantum channel can occur for any $\alpha<2$, also for measurements performed on single lattice sites. This result can be seen as complementary to the experimental observations in \cite{Richerme_etal14}, where supersonic propagation of correlations was observed for exponents up to $\alpha\approx1.19$ in a one-dimensional lattice.

\subsection{Density of states and typical propagation velocities}
\label{s:slowdown}

From figure~\ref{fig:kDependantCone} and the discussion in section~\ref{s:groupvelocity} we have seen that, while supersonic propagation can occur for $\alpha<2$, cone-like propagation is observed for these values of $\alpha$ at least for some initial states. In this section we will argue that the qualitative features of the observed behavior can be understood on the basis of the density of states
\begin{equation}
 \rho(v) = \frac{1}{2 \pi}\int^{2 \pi}_{0} \delta\left( v - \frac{\mathrm{d}\epsilon}{\mathrm{d}k} \right) \mathrm{d}k \label{eq:DOS01}
\end{equation}
in the large system limit. Eq.~(\ref{eq:DOS01}) can be rewritten as
\begin{equation}
\rho(v) = \frac{1}{2\pi} \sum_{k_0} \int^{2 \pi}_{0} \delta\left( k - k_0 \right) \left| \frac{\mathrm{d}^2}{\mathrm{d}k^2}\epsilon(k_0) \right|^{-1} \mathrm{d}k
\label{eq:Fourier_Identity}
\end{equation}
where the sum is taken over all roots $k_0$ of the argument of the delta function.  The polylogarithms that appear in the dispersion relation (\ref{eq:DispersionRelationBigSystem}) can be analytically evaluated for certain integer values of $\alpha$, yielding
\begin{eqnarray}
\rho(v)  &=&  \frac{1}{\pi} 
\cases{\displaystyle \frac{1}{1+v^2} & \text{for $\alpha=1$},\\
\displaystyle \frac{1}{2}\Theta\left( \pi -v \right) \Theta\left( \pi + v \right) & \text{for $\alpha=2$},\\
\displaystyle \frac{1}{\sqrt{4-v^2}} & \text{for $\alpha\to\infty$},}
\end{eqnarray}
where $\Theta$ is the Heaviside step function. For those three values of $\alpha$, the density of states is plotted in figure~\ref{fig:DOS} (right), but other cases can be evaluated numerically (not shown in the figure). Again, as for the group velocity in figure~\ref{fig:DispersionRelation} and the classical information capacity in \ref{s:AppChannel}, we find a threshold value of $\alpha=2$, as explained in the following.

For $\alpha<2$, the density of states $\rho$ is nonzero for all $v$, implying that propagation is not bounded by any finite maximum velocity. The maximum of $\rho$, however, is at $v=0$ for all $\alpha<2$, and this gives an indication that slow propagation with a small velocity is favored, although larger velocities do occur [as in figure~\ref{fig:kDependantCone} (left and center)]. The maximum at $v=0$ becomes more sharply peaked when $\alpha$ approached zero, explaining the vanishing of the inverse slope of the cone in figure~\ref{fig:kDependantCone} in that limit, as shown in figure~\ref{fig:DOS} (left).

For $\alpha\geq2$, the density of states $\rho$ is nonzero only on a finite interval $[-v_\mathrm{max},+v_\mathrm{max}]$, where $v_\mathrm{max}$ depends on $\alpha$. For $\alpha>2$ the density of states diverges, and therefore takes on its maximum, at $\pm v_\mathrm{max}$. This implies that the maximum velocity is favored, although smaller velocities also occur [as in figure~\ref{fig:kDependantCone} (right)].

\section{Conclusions}

In this paper we have studied, from several different perspectives, the nonequilibrium dynamics of lattice models with long-range interactions or long-range hopping, and in particular the propagation in space and time of correlations and other physical quantities. The focus of our work is on the competition between linear, cone-like propagation and faster-than-linear, supersonic propagation. We illustrate this competition in two regimes, both relevant for experimental realizations of long-range many-body systems in cold atoms, ions, or molecules:
\begin{itemize}
 \item[(i)] For small exponents $\alpha<2$ we prove that supersonic propagation can occur. At the same time, in such systems cone-like spreading can be the dominant form of propagation, with supersonic effects appearing only as small corrections [as in figure~\ref{fig:kDependantCone} (center)].
 \item[(ii)] For intermediate exponents (roughly between 3 and 8), propagation is observed to be linear initially, with supersonic effects setting in at larger times and distances [as in figure~\ref{f:horizon} (center) and figure~\ref{f:MExp} (right)].
\end{itemize}
To explain these observations, we provide model calculations as well as general bounds that provide a comprehensive and consistent picture of the various shapes of propagation fronts that can occur. Two of our results are Lieb-Robinson-type bounds, valid for large classes of models with long-range interactions. The first is a bound for models with exponents $\alpha$ smaller than the lattice dimension $D$, a regime for which hitherto no such bounds existed. Key to deriving the bound is the insight that for $\alpha<D$ the propagation speed in general scales asymptotically like a power law with the system size, and a meaningful bound therefore has to be derived in rescaled time $\tau$ as defined in (\ref{e:tau}). In physical time $t$, the bound then describes the increase of the propagation speed with increasing lattice size, as illustrated in figure~\ref{f:speedup}. The second Lieb-Robinson-type bound we report is essentially a cheat, as we stop half way through the derivation of a ``conventional'' Lieb-Robinson bound. Specializing this result to single-site observables and Hamiltonians with pair interactions only, we obtain an expression that can be evaluated numerically in an efficient way, easily reaching system sizes of $\mathscr{O}(10^4)$. This bound (\ref{e:MatrixExp}) is sharp enough to capture cone-like as well as supersonic behavior. In experimental studies of propagation in long-range interacting lattice models \cite{Richerme_etal14,Jurcevic_etal14}, the currently feasible lattice sizes are small and measured data can be compared to results from exact diagonalization. However, experimental work on systems of larger size is in progress, and exact diagonalization will not be feasible in that case. We expect that the matrix exponential bound (\ref{e:MatrixExp}) can provide guidance and sanity checks when analyzing the results of such experiments.

In the second half of the paper we complemented the bounds with results of one of the simplest long-range quantum models, namely a fermionic long-range hopping model in one dimension. We observed that cone-like propagation fronts can be a dominant feature also for small values of $\alpha$, and we explain the opening angle of such a cone, as well as the interplay of cone-like and supersonic features, on the basis of the dispersion relation combined with the density of states. These results indicate that it will depend crucially on the $k$-modes occupied whether cone-like or supersonic propagation is dominant. We expect that such an improved understanding can provide guidance for optimizing experimental efforts to harness long-range interactions in a variety of quantum information and signaling tasks.

\appendix

\section{Lieb-Robinson bounds}

\subsection{Derivation of the matrix exponential bound}
\label{s:AppMatrixExp}

As in section~\ref{s:smallalpha}, we consider a $D$-dimensional lattice $\Lambda$ consisting of $N$ sites, a Hilbert space (\ref{e:Hilbert}) consisting of finite-dimensional local Hilbert spaces, and a generic Hamiltonian with pair interactions (\ref{e:Hpair}). Let $O_A$ and $O_B$ be two bounded linear operators compactly supported on subsets $A,B\subset\Lambda$ with $A\cap B=\emptyset$. Under these conditions, similar to the derivation of Eq.~(2.12) of Ref.~\cite{NachtergaeleOgataSims06}, one can derive the upper bound
\begin{equation}\label{e:LRBsupp}
\left|\left|\left[O_A(t),O_B(0)\right]\right|\right|
\leq 2 \left|\left| O_A\right|\right| \left|\left| O_B\right|\right| \sum_{n=1}^\infty \frac{(2 \left|t\right|)^n}{n!}a_n
\end{equation}
For pair interactions, and considering observables $O_A$ and $O_B$ supported on single lattice sites only (i.e., $A=\{i\}$ and $B=\{j\}$), the coefficients $a_n$ are upper bounded by
\begin{equation}\label{e:CoefficientsSingleSiteSupport}
 a_n\leq \kappa^n \sum_{k_1,\dots,k_{n-1}} J_{i,k_1}J_{k_1,k_2}\dots J_{k_{n-1},j} =\kappa^n\left(J^n\right)_{i,j},
\end{equation}
where $J$ is the interaction matrix with elements $J_{k,l}=\left|\left| h_{kl}\right|\right|$ and $\kappa = \sup_{q\in \Lambda} \sum_k J_{q,k}$. Then (\ref{e:LRBsupp}) can be written as
\begin{equation}
 \label{e:LRBMatrixExpsupp}
 \frac{\left|\left|\left[O_A(t),O_B(0)\right]\right|\right|}{2 \left|\left| O_A\right|\right| \left|\left| O_B\right|\right|}
\leq \left(\sum_{n=1}^\infty \frac{(2\kappa \left|t\right|)^n}{n!}J^n\right)_{i,j}= \exp\left(2 J\kappa \left|t\right|\right)_{i,j} - \delta_{i,j},
\end{equation}
which proves (\ref{e:MatrixExp}).

For translationally invariant one-dimensional lattices, $J$ is a circulant matrix and can be diagonalized by means of Fourier transformation. For the example of power law interactions $h_{kl}\propto\dist(k,l)^{-\alpha}$, the diagonal elements of the Fourier-transformed matrix $J$ are given by
\begin{equation}\label{e:epsilon_k}
 \epsilon(k)=\sum_{n=1}^{\lfloor (N-1)/2 \rfloor} \frac{\cos(nk)}{n^\alpha} \underbrace{\color{gray}+\frac{\ee^{ik N/2}}{2(N/2)^\alpha}}_{\mathrm{if}\ N\ \mathrm{even}}
\end{equation}
with $k=2\pi m/N$, $0\leq m < N$. Using these eigenvalues, $J$ can be exponentiated in the diagonal basis, followed by an inverse Fourier transformation to evaluate the Lieb-Robinson bound (\ref{e:MatrixExp}).

We envisage the bound (\ref{e:MatrixExp}) to be particularly useful for finite systems of intermediate size where the matrix exponential can be computed numerically with reasonable effort. However, since (\ref{e:MatrixExp}) is sharper than the bounds in \cite{HastingsKoma06,NachtergaeleOgataSims06}, a thermodynamic limit will exist (at least) under the same conditions required in those proofs, and in particular for $D$-dimensional regular lattices with power law interactions and exponents $\alpha>D$.

\subsection{Lieb-Robinson bounds in rescaled time} 
\label{s:AppRescaledBound}
As in section~\ref{s:MatrixExp}, we consider a $D$-dimensional lattice $\Lambda$ consisting of $N$ sites, a Hilbert space (\ref{e:Hilbert}) with finite-dimensional local Hilbert spaces, and a generic Hamiltonian $H$ with $n$-body interactions (\ref{e:Hgeneric}). We require that $H$ satisfies conditions (\ref{e:Cond1}) and (\ref{e:Cond2}). For the proof of a Lieb-Robinson-type bound, we follow the general strategy of \cite{NachtergaeleOgataSims06}, augmented with the $\mathscr{N}_\Lambda$-rescaling taken from \cite{MetivierBachelardKastner14}.

As a shorthand we introduce
\begin{equation}\label{eqn:ChiTau}
 \chi(t)=\left[O_A(t),O_B \right].
\end{equation}
for the commutator on the left-hand side of (\ref{e:LRBRescaledTime}). Differentiating with respect to $t$ yields
\begin{equation}\label{eqn:ChiTauPrime}
 \chi^\prime(t)=\mathrm{i} \left[\chi(t),I_A(t) \right] + \mathrm{i} \left[ O_A(t),\left[I_A(t),O_B\right]\right],
\end{equation}
where $I_A=\sum_{Z:Z\cap A\neq \emptyset}h_Z$ is the set of local Hamiltonian terms that have non-zero overlap with $A$. Using the boundedness of $O_A(t)$ we apply Lemma A.1 of Ref.~\cite{NachtergaeleOgataSims06} to the norm-preserving first term of (\ref{eqn:ChiTauPrime}), yielding 
\begin{equation}\label{eqn:LRIneq}
 \left|\left| \chi(t) \right|\right|-\left|\left| \chi(0) \right|\right| \leq 2\left|\left| O_A\right|\right| \sum_{Z:Z\cap A\neq \emptyset}\int_0^{|t|} \left|\left| \left[h_Z(s),O_B \right]\right|\right| \mathrm{d}s.
\end{equation}
Next we define
\begin{equation}
 C_{O_B}(A,t):=\sup_{O_A\in \mathcal{O}_A} \frac{\left|\left|\chi(t)\right|\right|}{\left|\left| O_A \right|\right|},
\end{equation}
where $\mathcal{O}_A$ is the set of observables compactly supported on $A$. Making use of this definition, (\ref{eqn:LRIneq}) can be rewritten as
\begin{equation}\label{eqn:IneqIterate}
 \frac{C_{O_B}(A,t)-C_{O_B}(0,t)}{2}\leq \sum_{Z:Z\cap A\neq \emptyset}\int_0^{|t|}\left|\left| h_Z\right|\right| C_{O_B}(Z,s) \mathrm{d}s.
\end{equation}
Eq.~(\ref{eqn:IneqIterate}) can be applied recursively to show that
\begin{equation}\label{eqn:CBTauSeries}
 C_{O_B}(A,t)\leq 2\left|\left| O_B \right|\right| \sum_{n=1}^\infty \frac{(2\left|t\right|)^n}{n!}a_n
\end{equation}
with coefficients
\begin{equation}\label{e:an}
 a_n=\sum_{{Z_1\subset\Lambda \atop Z_1\cap A\neq \emptyset}} \sum_{{Z_2\subset\Lambda\atop Z_2\cap Z_1\neq \emptyset}}\dots \sum_{{Z_n\subset\Lambda\atop Z_n\cap Z_{n-1}\neq \emptyset}}   \prod_{l=1}^n \ \left|\left| h_{Z_l}\right|\right| \delta_B(Z_n),
\end{equation}
where
\begin{eqnarray}
 \delta_B(Z)  &=& \cases{\displaystyle0 & \text{if $Z\cap B\neq \emptyset$},\\
\displaystyle 1 & \text{otherwise}.}
\end{eqnarray}
Under the conditions (\ref{e:Cond1}) and (\ref{e:Cond2}) these coefficients can be bounded by
\begin{equation}\label{e:anbound}
 a_n\leq   \frac{p^{n-1}\lambda^n}{\mathscr{N}_\Lambda^n(1+\dist(A,B))^\alpha}.
\end{equation}
Inserting (\ref{e:anbound}) into (\ref{eqn:CBTauSeries}) and using the definition (\ref{e:tau}) of rescaled time $\tau$, one obtains
\begin{equation}
 C_{O_B}(A,t)\leq \frac{2\left|\left| O_B \right|\right| \left|A\right|\left|B\right|}{p(1+\dist(A,B))^\alpha} \left(\exp\left[2p \lambda \left|\tau\right|\right]-1 \right),
\end{equation}
and this implies the bound
\begin{equation}\label{e:LRBRescaledTimeAppendix}
 \left|\left| \left[ O_A(\tau\mathscr{N}_\Lambda),O_B \right] \right|\right| \leq \frac{2\left|\left| O_A \right|\right|\left|\left| O_B \right|\right| \left|A\right|\left|B\right|}{p(1+\dist(A,B))^\alpha} \left(\exp\left[2p \lambda \left|\tau\right| \right]-1 \right)
\end{equation}
in rescaled time $\tau$, valid for power law interactions with exponents $\alpha > 0$.

\subsection{Discussion of the bound in Reference \texorpdfstring{\cite{GongFossFeigMichalakisGorshkov14}}{[27]}} 
\label{s:AppGong}

In \cite{GongFossFeigMichalakisGorshkov14} a Lieb-Robinson-type bound was derived whose functional form consists of a linear (cone-like) and a faster-than-linear (supersonic) contribution. This bound is a major improvement over that in \cite{HastingsKoma06} in the regime of large $\alpha$, where the former becomes more and more similar to a nearest-neighbour bound, as it should. Here we scrutinize the applicability of the bound in \cite{GongFossFeigMichalakisGorshkov14} for describing the cone-like and supersonic features of long-range models with intermediate exponents $\alpha$ (roughly in the range $3\leq\alpha\leq8$).

The bound in \cite{GongFossFeigMichalakisGorshkov14} is derived for Hamiltonians
\begin{equation}
H=\frac{1}{2}\sum_{i\neq j}h_{ij}
\end{equation}
with two-body interactions $ h_{ij}$ satisfying
\begin{equation}
\left|\left| h_{ij}\right|\right| \leq\frac{1}{\dist(i,j)^\alpha}
\end{equation}
on $D$-dimensional regular cubic lattices. For exponents $\alpha\geq1$ a bound of the form
\begin{equation}\label{e:bound}
\frac{\left|\left| [A(t),B]\right|\right|}{2\left| A\right| \left| B\right|} \leq T_1 + T_2
\end{equation}
is obtained, where $A$ and $B$ are observables on lattice sites that are a distance $\delta$ apart, and
\begin{equation}\label{e:T1T2}
T_1 = c_1 \frac{e^{v_1 t}-1}{e^{\mu \delta}},\qquad T_2 = c_2 \frac{e^{v_2 t}-1}{[(1-\mu)\delta]^\alpha},
\end{equation}
with $c_1=\lambda^{-1}$, $v_1=2\lambda^2 e$, $c_2=(\lambda9^D)^{-1}$, $v_2=2\lambda^2 9^D$, $\lambda=\sum_k d(i,k)^{-\alpha}$, and $0<\mu<1$. $T_1$ has the same functional form as the classic Lieb-Robinson bound for Hamiltonians with finite-range interactions \cite{LiebRobinson72}, which is known to produce a linear, cone-shaped causal region. $T_2$ has the functional form of the bound originally derived by Hastings and Koma \cite{HastingsKoma06}. Both, $T_1$ and $T_2$ contain the free parameter $\mu$, which determines, among other things, the slope of the linear soundcone. So the ``velocity'' associated with the cone can be tuned to an arbitrary value, irrespectively of the physical behavior of the model studied.

\begin{figure}[t]
{\center 
\includegraphics[height=0.3\linewidth]{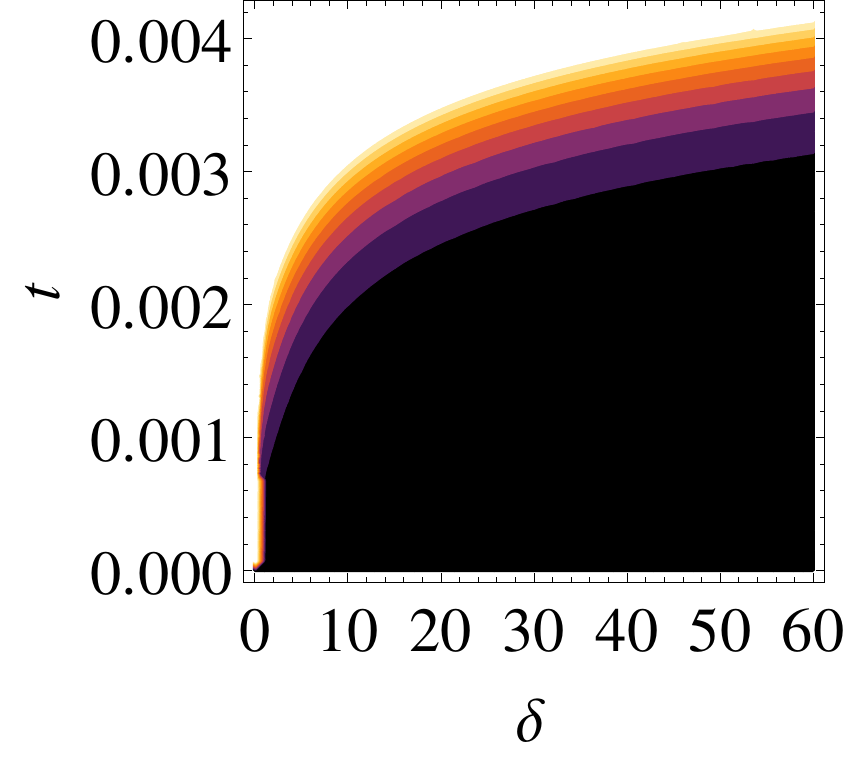}
\hfill
\includegraphics[height=0.3\linewidth]{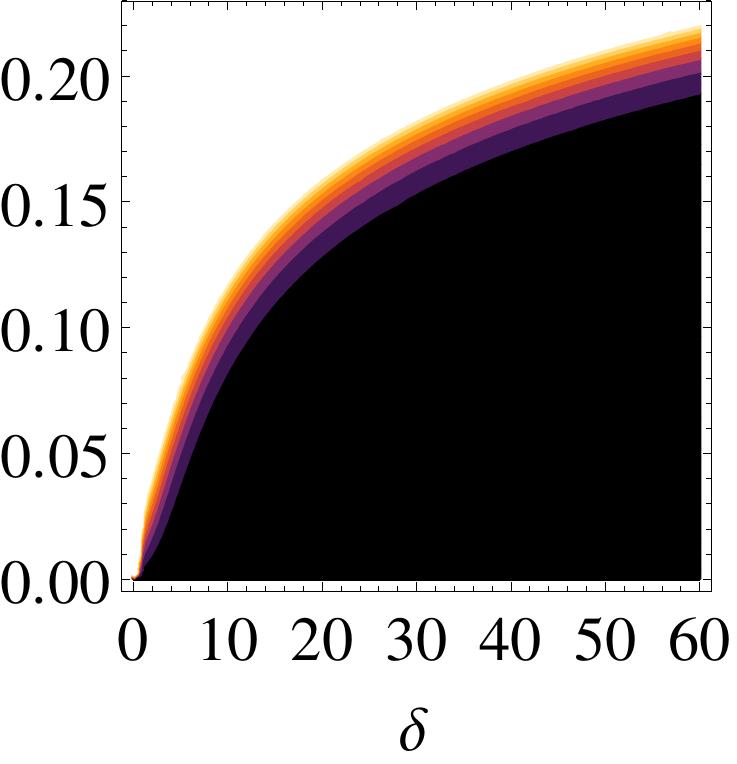}
\hfill
\includegraphics[height=0.3\linewidth]{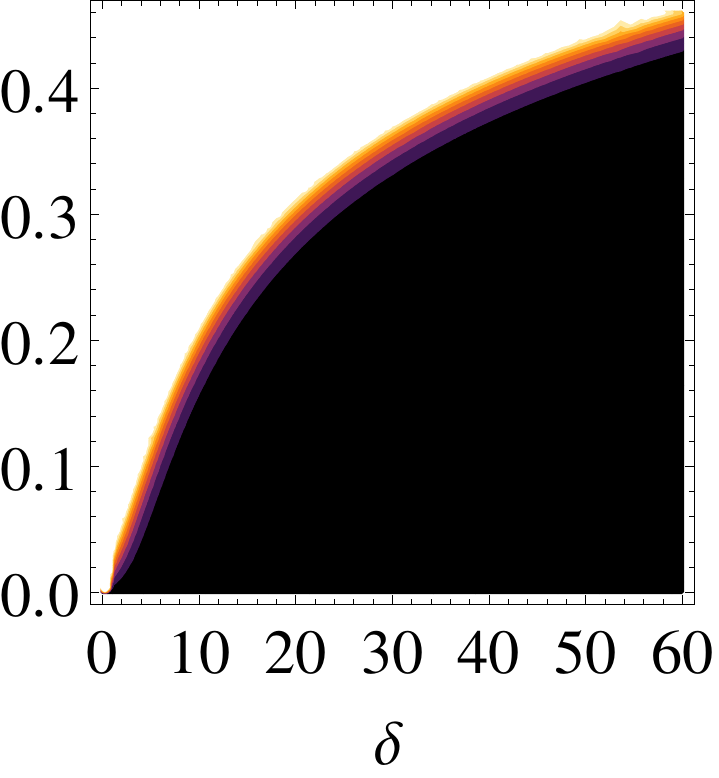}
\hfill
\includegraphics[height=0.3\linewidth]{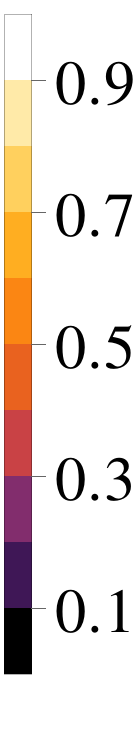}
}
\caption{\label{f:contours}%
Contour plots of the bound (\ref{e:optimised}) in one spatial dimension for $\alpha=6/5$ (left), $4$ (center), and $8$ (right). Even for $\alpha=8$ there is at best a hint of a linear regime. See the matrix exponential bound in figure~\ref{f:MExp} for comparison.}
\end{figure}

Based on (\ref{e:bound}) and (\ref{e:T1T2}), the sharpest bound
\begin{equation}\label{e:optimised}
B(\delta,t)=\min_\mu\left(c_1 \frac{e^{v_1 t}-1}{e^{\mu \delta}} +  c_2 \frac{e^{v_2 t}-1}{[(1-\mu)\delta]^\alpha}\right)
\end{equation}
is obtained by minimizing, separately for each value of $\delta$ and $t$, over the free parameter $\mu$. From the contour plots of $B$ in figure~\ref{f:contours} it becomes clear that the ``linearity'' of $T_1$ can be deceiving, as a linear, cone-like regime is not particularly prominent, not even for larger exponents like $\alpha=8$. Of course it is always possible to construct a linear-looking bound by weakening $B$, but this would be unrelated to the physical behavior of the class of models studied. 

Another, more sophisticated bound has recently been put forward in \cite{FossFeigGongClarkGorshkov15}, but the form of the propagation front has not yet been analyzed and discussed (beyond the long-distance asymptotics).

\section{Information capacity of the long-range Ising model}
\label{s:AppChannel}

In this appendix we prove that supersonic transmission through a quantum channel can occur for any $\alpha<2$, also for measurements performed on single lattice sites. Like for the study of the group velocity of the long-range hopping model in section~\ref{s:groupvelocity}, we find a threshold value of $\alpha=2$ below which propagation becomes supersonic. The proof uses techniques from Ref.~\cite{EisertvdWormManmanaKastner13} and applies them to a slightly more involved model for which supersonic propagation is found to occur also for single-site measurements.

We consider a finite one-dimensional lattice $\Lambda=\{1,\dots,N\}$ consisting of $N$ sites. To implement a quantum channel, we encode a signal on site 1, and measure the effect of that encoding after a time time at site $N$. On this lattice we define an Ising Hamiltonian with arbitrary couplings,
\begin{equation}
 H=\sum_{i<j} J_{ij} \sigma^z_i \sigma^z_j.
\end{equation}
Defining the sublattices $A=\{1\}$, $B=\{N\}$, and $S=\Lambda\setminus(A\cup B)$, the Hamiltonian can be rewritten as
\begin{equation}
 H = H_{AS} + H_{AB} + H_{SB} + H_{SS}
\end{equation}
with
\begin{equation}
 H_{XY} := \sum_{i\in X} \sum_{j \in Y} J_{ij}\sigma^z_i \sigma^z_j,
\end{equation}
where $X,Y\in\{A,S,B\}$. As an initial state we choose
\begin{equation}
 \rho(0) = |\downarrow\rangle_1 \, _1\langle{ \downarrow} | \bigotimes_{s\in S} |\downarrow \rangle_s \, _s\langle{\downarrow}|\otimes 
 |+\rangle_N \, _N \langle{ + }|
\end{equation}
with $\sigma_i^z\ket{\uparrow }_i=\ket{\uparrow }_i$, $\sigma_i^z\ket{\downarrow }_i=-\ket{\downarrow }_i$, and
$\ket{+}_j = \bigl( \ket{\uparrow}_{j} + \ket{ \downarrow }_{j} \bigr)/\sqrt{2}$.
Initially all the spins are pointing down, except the one at $B=\{N\}$. 

A binary quantum channel is implemented by starting the time evolution either with $\rho(0)$ (sending a ``0''), or starting with $U_{A} \rho(0) U^{\dagger}_{A}$ (sending a ``1''), where $U_{A}$ is a unitary supported on $A$ only. The classical information capacity $C_t$ can be bounded from below by the probability to detect, by measuring according to a positive operator valued measure $\pi_B$, a signal at $B$ after a time $t$, 
\begin{equation}\label{e:channelcapacity}
C_t \geq p_t=\bigl| \Tr\left\{ N_t\left[ \rho(0) \right] \pi_B \right\} - \Tr\left\{ T_t\left[ \rho(0)
\right] \pi_B\right\} \bigr|,
\end{equation}
with
\begin{eqnarray}
 N_t\left[ \rho(0) \right] :=& \mathrm{Tr}_{\Lambda\backslash B}\Bigl[ \ee^{-\ii Ht} \rho(0) \ee^{\ii Ht} \Bigr], \label{e:Nt}\\
 T_t\left[ \rho(0) \right] :=& \Tr_{\Lambda\backslash B}\Bigl[ \ee^{-\ii Ht} U_{A} \rho(0) U^{\dagger}_{A} \ee^{\ii Ht}\Bigr].
\end{eqnarray}
In the following we compute a lower bound on the right-hand side of (\ref{e:channelcapacity}), and study this bound as a function of the channel length, i.e., the distance between $A$ and $B$.  

We choose $\pi_B = | +\rangle_N\,_N \langle +|$ and $U_A=| \uparrow \rangle_1\,_1 \langle \downarrow |$, where the latter is a spin flip operator on the first lattice site. For the time-evolved density operator in (\ref{e:Nt}) we find
\begin{eqnarray}
\!\!\!\!\!\!\!\!\!\!\!\!\rho(t)&=
 \ee^{-\ii H_{AB}t} \ee^{-\ii H_{AS}t} \ee^{-\ii H_{SS}t} \ee^{-\ii H_{SB}t}\label{eq:trick}\\
&\quad \times \Bigg[ \bigotimes_{s=1}^{N-1} |\downarrow\rangle_s\,_s \langle\downarrow | \otimes | + \rangle_N\,_N \langle + | \Bigg] \ee^{\ii H_{SB}t} \ee^{\ii H_{SS}t} \ee^{\ii H_{AS}t}
\ee^{\ii H_{AB}t}\nonumber\\
&= \bigotimes_{s=1}^{N-1} |\downarrow\rangle_s\,_s \langle\downarrow | \left\{ \exp\left[\ii t \sum_{r=1}^{N-1} J_{rN}  \sigma^z_N \right] |+\rangle_N\,_N \langle + | \exp\left[-\ii t\sum_{r=1}^{N-1} J_{rN} \sigma^z_N
 \right]  \right\}.\nonumber
\end{eqnarray}
All the exponentials not supported on $B$ add up to zero since the initial state prepared on $\Lambda\backslash B$ is an eigenstate of the Ising Hamiltonian.  Taking the trace gives 
\begin{equation}
\Tr\left\{ N_t \left[ \rho(0) \right] \pi_B \right\} = 
 \frac{1}{2}\left\{ 1 + \cos\left[ 2t \left(\sum_{r\in S} J_{rN} + J_{1N} \right) \right] \right\} .
\end{equation}
A similar calculation shows that
\begin{equation}
 \Tr\left\{ T_t\left[ \rho(0) \right] \pi_B \right\} = \frac{1}{2}\left\{ 1 + \cos\left[ 2t
\left(\sum_{r\in S} J_{rN} - J_{1N} \right) \right] \right\}.
\end{equation}
The probability of detecting a signal in $B$ at some time $t>0$ is then given by
\begin{eqnarray}
 p_t&= \frac{1}{2}\left|\cos\left[ 2t \left( \sum_{r\in
S} J_{rN} + J_{1N} \right)\right] - \cos\left[ 2t \left( \sum_{r\in S} J_{rN} - J_{1N} \right)\right]\right| \\
&=
\left| \sin\left(2 t \sum_{r\in S} J_{rN} \right) \sin\left( 2tJ_{1N} \right) \right|.
\label{e:pt}
\end{eqnarray}
To derive a nontrivial (nonzero) lower bound on $p_t$, we target the regime before oscillatory behavior in (\ref{e:pt}) sets in. Using the inequality
\begin{equation}
\sin x\geq \frac{2x}{\pi}\qquad\mathrm{for}\ 0\leq x\leq\pi/2
\end{equation}
and assuming power law interactions $J_{ij}=|i-j|^{-\alpha}$, we obtain
\begin{equation}\label{e:pt2}
p_t\geq \frac{4t}{\pi}\frac{1}{(N-1)^\alpha}\frac{4t}{\pi}\sum_{r=2}^{N-1}\frac{1}{(N-r)^\alpha},
\end{equation}
valid for times
\begin{equation}
t\leq\frac{\pi}{4}\bigg/\sum_{r=1}^{N-2}\frac{1}{r^\alpha}.
\end{equation}
Interpreting the sum in (\ref{e:pt2}) as an upper Riemann sum, we have
\begin{equation}
\sum_{r=2}^{N-1}\frac{1}{(N-r)^\alpha} = \sum_{r=1}^{N-2}\frac{1}{r^\alpha} > \int_0^{N-2} \frac{\dd r}{(r+1)^\alpha}.
\end{equation}
Then we can bound $p_t$ by
\begin{equation}\label{e:finalbound}
p_t> \frac{16t^2}{\pi^2(\alpha-1)}\frac{1}{(N-1)^\alpha}\left(1-\frac{1}{(N-1)^{\alpha-1}}\right)=:\underline{p}_t.
\end{equation}
For $\alpha>1$ and large $N$ the second term in the square bracket in (\ref{e:finalbound}) is much smaller than 1, and we obtain
\begin{equation}\label{e:boundasymptotic}
\underline{p}_t \sim \frac{16t^2}{\pi^2(\alpha-1)}\frac{1}{(N-1)^\alpha}
\end{equation}
for the large-$N$ asymptotic behavior of the bound $\underline{p}_t$. In our setting, $\delta=N-1$ is the distance between the regions $A$ and $B$. To determine the shape of a contour line at which $\underline{p}_t$ is equal to some constant $\epsilon$, we set
\begin{equation}
\epsilon=\underline{p}_t\propto \frac{t^2}{\delta^\alpha},
\end{equation}
and we can read off that
\begin{equation}\label{e:deltapropt}
 \delta \propto t^{2/\alpha}
\end{equation}
along any of those contour lines. Eq.~(\ref{e:deltapropt}) describes faster-than-linear (supersonic) growth of $\delta$ for $\alpha<2$. It is straightforward to extend the above calculation to more general initial conditions as well as to lattices of arbitrary dimension.

\ack
\addcontentsline{toc}{section}{Acknowledgments}
The authors acknowledge helpful discussions with Jens Eisert, Fabian Essler, Michael Foss-Feig, Alexey Gorshkov, Stefan Kehrein, Salvatore Manmana, Ryan Sweke and Davide Vodola. D.\,S.\ acknowledges financial support by the Studienstiftung des deutschen Volkes; M.\,K. by the National Research Foundation of South Africa through the Incentive Funding and the Competitive Programme for Rated Researchers.\\


\section*{References}
\addcontentsline{toc}{section}{References}
\bibliographystyle{iopart-num}
\bibliography{LRLR}

\end{document}